\documentclass[aps,prl,twocolumn,superscriptaddress,nobibnotes,nolongbibliography]{revtex4-1}

\usepackage[T1]{fontenc}
\usepackage[utf8]{inputenc}
\usepackage[english]{babel}

\usepackage{amsmath,amssymb,mathrsfs}
\usepackage{graphicx}
\usepackage{xcolor}
\usepackage{hyperref}
\usepackage{nicefrac}

\hypersetup{
    colorlinks=true,
    citecolor=blue,
    linkcolor=blue,
    urlcolor=blue
}

\usepackage[final]{pdfpages}

\makeatletter
\AtBeginDocument{\let\LS@rot\@undefined}
\makeatother
\begin{document}
      
\title{Two-mode geometry controls multiscale organization in bipartite systems}
\author{Ottavia Falconi}
\affiliation{Physics Department, University of Rome Tor Vergata, 00133 Rome, Italy}
\affiliation{Enrico Fermi Research Center (CREF), 00184, Rome, Italy}
\author{Giulio Cimini}
\affiliation{Physics Department and INFN, University of Rome Tor Vergata, 00133 Rome, Italy}
\affiliation{Enrico Fermi Research Center (CREF), 00184, Rome, Italy}
\author{Pablo Villegas}
\email{pablo.villegas@cref.it}
\affiliation{Enrico Fermi Research Center (CREF), 00184, Rome, Italy}
\affiliation{Instituto Carlos I de F\'isica Te\'orica y Computacional, Universidad de Granada, 18071 Granada, Spain}

\begin{abstract}
Many complex systems are organized around complementary roles and naturally described as bipartite networks. Unveiling their multiscale structure presents a fundamental challenge because coarse-graining procedures must preserve role separation, whereas standard approaches collapse it via one-mode projections. Here we introduce a Laplacian-based renormalization framework that operates directly on the bipartite architecture, enabling scale transformations while retaining role differentiation.  Using controlled bipartite ensembles at criticality, we show that structural imbalance systematically reshapes organization across scales while leaving scaling properties invariant, revealing a separation between universality and geometry. Applying the coarse-graining framework to empirical bipartite networks, we uncover nontrivial multiscale hierarchies for both roles. 
In contrast, renormalization performed after one-mode projection—which truncates diffusion paths to nearest neighbors—yields qualitatively different structures. 
Our results identify two-mode geometry as a fundamental constraint for revealing multiscale organization in systems with role separation.
\end{abstract}
\maketitle 

Organization in complex systems often arises from structured differences \cite{Simon1962}, with many systems composed of distinct classes whose identities are defined through their mutual relations \cite{Bascompte2003,Latapy2008,LatoraBook}. 
Bipartite networks provide a canonical representation of such role-differentiated systems, encoding interactions exclusively across disjoint node classes \cite{Borgatti1997,Holme2003,Guillaume2004}. 
From ecological host–parasite interactions to social affiliation networks \cite{Bastolla2009,Hidalgo2007,Jeong2000,Newman2001}, the absence of intra-class links is not incidental but constitutive: it constrains connectivity to alternating paths across classes, embedding role separation directly into the geometry of interaction pathways.

This constraint shapes organization across scales. In ecological systems such as plant–pollinator or host–parasite networks, interactions form nested and modular patterns, where tightly interacting groups are embedded within larger functional or phylogenetic layers \cite{Bascompte2007}. Such multiscale organization underpins key system-level properties, including robustness, biodiversity, and the propagation of perturbations, as species loss can cascade differently within and across modules \cite{Bastolla2009,Guimaraes2020}. 

Capturing these hierarchies requires coarse-graining methods that preserve the underlying bipartite architecture.
Current approaches, however, typically neglect this constraint. Bipartite community detection methods operate at a single resolution \cite{Barber2007,Guimera2007,Larremore2014,Fortunato2010,Peixoto2014}, while multiscale structure is often inferred from one-mode projections \cite{Zhou2007,Newman2001}, in which nodes in the same class are connected based on shared neighbors in the other class. Projection is a lossy representation \cite{Kitsak2017} that 
reshapes the geometry of interactions. 
As a result, structural information may already be distorted before any explicit coarse-graining is performed.

This raises a fundamental question: is multiscale organization in bipartite systems invariant under representation, or does structural projection alter the geometry from which hierarchy emerges? In other words, can the organization of bipartite systems be recovered from one-mode projections, or is it an intrinsic property of the two-mode architecture? 

Addressing this question requires a renormalization framework that operates directly on bipartite networks. 
Diffusion-based renormalization approaches have recently shown that multiscale organization in complex heterogeneous networks can be uncovered by integrating Laplacian diffusion modes \cite{LRG,Modularity,InfoCore,JSTAT,PRL_Poggialini}, yielding coarse-grained representations that preserve intrinsic information flow. However, these formulations are designed for monopartite systems and do not preserve role separation when applied to bipartite structures.

Here we develop a Laplacian-based renormalization framework that preserves the two-mode geometry across scales. 
Applied to random bipartite networks at criticality, our approach shows that structural imbalance reorganizes connectivity without affecting scaling behavior. 
When applied to empirical systems, it uncovers multiscale hierarchies that remain hidden under one-mode projections, where diffusion is effectively restricted to short paths.
Together, these results demonstrate that multiscale organization depends on the underlying representation, and that altering the two-mode geometry can prevent its faithful reconstruction. 

\begin{figure*}
    \centering
    \includegraphics[width=0.95\textwidth]{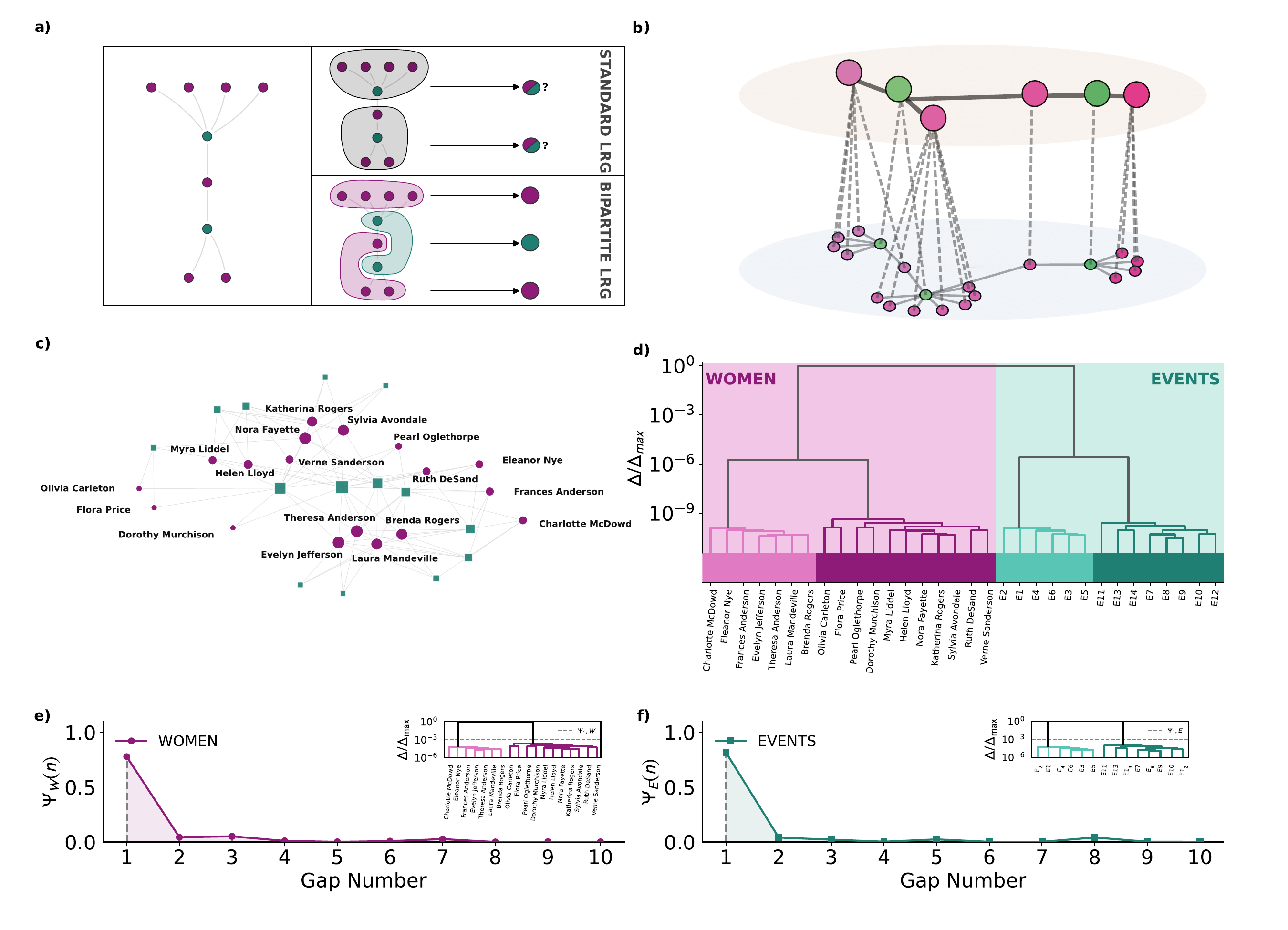}
   \caption{\textbf{Two-mode-preserving renormalization in bipartite networks.}
\textbf{(a)}, Diffusion-driven coarse-graining in a bipartite architecture. Node classes are identified by different colors. While standard Laplacian renormalization creates mixed blocks, the bipartite renormalization preserves separation between classes. 
\textbf{(b)}, Illustration of a coarse-graining step of the bipartite renormalization scheme, where the scale is increased by integrating fast Laplacian modes. Macronodes are formed independently within each partition, enforcing the absence of intra-class connectivity at all scales. 
\textbf{(c)}, Davis Southern Women network, a prototypical bipartite system of individuals (women) and events they participated to. 
\textbf{(d)}, Diffusion-based dendrogram of the network at fixed timescale $\tau$, showing a hierarchical organization consistent with the known bipartite modular structure. 
\textbf{(e,f)}, Gap function $\Psi(n)$ evaluated separately on the two partitions (women and events). The first non-trivial maximum ($n=1$) identifies the leading intra-partition branching beyond the trivial bipartite split, indicating robust mesoscale structure.}
    \label{fig:panel_1}
\end{figure*}

\subsection*{Laplacian renormalization of bipartite networks}

Because bipartite interactions are restricted across disjoint node classes, any coarse-graining procedure respecting this constraint must define supernodes within each class, ensuring the absence of intra-class links at all scales. Merging nodes across classes would instead generate effective intra-class connectivity and alter the underlying diffusion geometry.

Our approach builds on diffusion dynamics governed by the graph Laplacian $\hat L$, whose propagator $e^{-\tau \hat L}$ encodes communication between nodes across all paths at a given timescale $\tau$. From this propagator, we define the Laplacian density matrix \cite{Domenico2016,InfoCore}, $\hat \rho(\tau) = \frac{e^{-\tau \hat L}}{\mathrm{Tr}\, (e^{-\tau \hat L})}$,
which provides a normalized measure of diffusion-based communicability. Following this framework, we define an effective distance between nodes \cite{Modularity},
$\mathcal{D}_{ij}(\tau) = \frac{1 - \delta_{ij}}{\rho_{ij}(\tau)}$, so that nodes with stronger diffusion-mediated communication are closer. This distance induces a hierarchical geometry on the network, allowing the construction of a dendrogram at fixed diffusion time $\tau$ via standard hierarchical clustering (using standard linkage methods), as in the monopartite LRG framework \cite{LRG,JSTAT}. 
Coarse-graining is then performed by cutting the dendrogram at a chosen resolution and merging nodes within each cluster into supernodes \cite{LRG}. Varying $\tau$ effectively scans the network across scales, progressively integrating fast diffusion modes and yielding a multiscale hierarchy.
In the following, we focus on the small-$\tau$ regime, where all Laplacian modes contribute to the propagator, ensuring that the induced geometry captures the full multiscale structure of the network \cite{Modularity}.

However, when directly applied to bipartite systems, this procedure can merge nodes across classes, generating effective intra-class connectivity and altering the diffusion geometry. To preserve the bipartite constraint, we impose an infinite geometric penalty on inter-class pairs:
\[
\mathcal{D}^{B}_{ij}(\tau)
=
\frac{1 - \delta_{ij}}{\rho_{ij}(\tau)} \, \mathbf{1}_{\mathrm{intra\text{-}class}}
+
\frac{1}{\varepsilon} \, \mathbf{1}_{\mathrm{inter\text{-}class}},
\qquad \varepsilon \to 0^{+}.
\]
In this limit, clustering is restricted to nodes within the same class, yielding a constrained Kadanoff blocking scheme that preserves the bipartite architecture at all scales (Fig.~\ref{fig:panel_1}a,b).
The resulting coarse-grained network inherits inter-class connectivity from the original graph: two supernodes are connected if at least one cross-class link exists between their constituent nodes, while no intra-class edges are introduced.

We illustrate application of the bipartite Laplacian renormalization group (b-LRG) on the Davis Southern Women network (Fig.~\ref{fig:panel_1}c), a prototypical bipartite social system~\cite{Davis1941}. The diffusion-based dendrogram at fixed $\tau$, obtained from the distance $\mathcal{D}^B(\tau)$, recovers the two primary modules reported in the literature—corresponding to two groups of women attending distinct sets of events \cite{Barber2007,Cui2014} (Fig.~\ref{fig:panel_1}d).
To quantify the statistical significance of branching across scales, we evaluate for each class the partition stability function $\Psi$ \cite{Modularity}, defined as the relative separation between consecutive dendrogram levels (see Methods). The first non-trivial maximum identifies robust intra-partition structure beyond the trivial bipartite split, revealing hierarchical organization while preserving role differentiation (Fig.~\ref{fig:panel_1}e,f).

\subsection*{Universality and geometry in bipartite networks}

To illustrate how the bipartite structure shapes multiscale organization, we study random bipartite ensembles at criticality (see Methods), where scale invariance provides a controlled benchmark. In these systems, the percolation threshold depends 
on the average degree $\langle k\rangle=2E/(N_A+N_B)$ as well as on the partition asymmetry $\alpha = N_A/N_B$, where $N_A$ and $N_B$ are the number of nodes in the two classes $A$ and $B$, and $E$ is the number of edges. Varying the imbalance while adjusting the average degree allows us to probe different regimes while remaining at criticality.

\begin{figure}
    \centering
    \includegraphics[width=1.0\columnwidth]{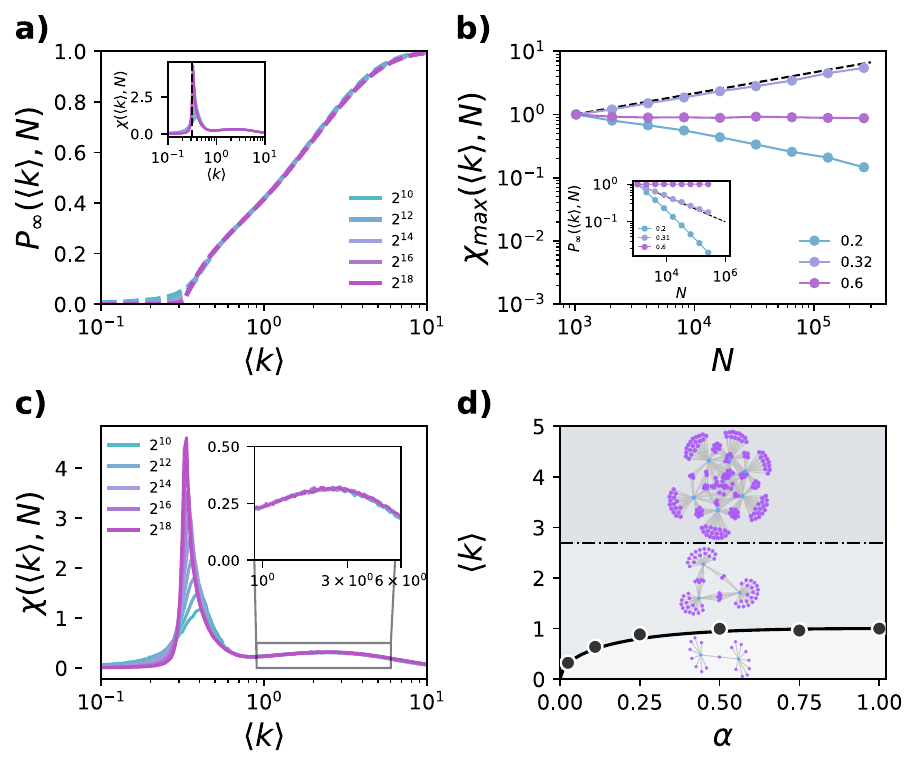}
    \caption{\textbf{Critical scaling and geometry in random bipartite networks.}
\textbf{(a)} Giant-component size $P_\infty$ as a function of the average degree $\langle k \rangle$ (computed on the full bipartite network) for a strongly imbalanced system ($\alpha=0.025$) and increasing system size $N=N_A+N_B$.  
\textbf{(b)} Finite-size scaling at criticality for different imbalance values $\alpha$. The maximum susceptibility scales as $\chi_{\max}(k_c,N)\sim N^{\gamma}$ (main panel), while the order parameter scales as $P_\infty(k_c,N)\sim N^{-\beta}$ (inset). Dashed lines indicate the mean-field percolation exponents $\beta=\gamma=1/3$. 
\textbf{(c)} Susceptibility $\chi(\langle k \rangle,N)$ as a function of $\langle k \rangle$ for increasing system size. Besides the main critical peak, a broader secondary feature emerges at larger $\langle k \rangle$ (inset), indicating an additional structural scale that becomes visible under strong imbalance. 
\textbf{(d)} Phase diagram in the $(\alpha,\langle k\rangle)$ plane. The solid curve denotes the analytical critical line $k_c(\alpha)=2\sqrt{\alpha}/(1+\alpha)$ separating the fragmented and percolating regimes; symbols mark the numerical estimates of the transition. The dashed-dotted line marks the localization transition corresponding to the second peak of $\chi$. Snapshots illustrate representative network geometries across imbalance values.}
    \label{fig:scaleInv}
\end{figure}

We first control how critical scaling depends on class imbalance. The giant-component size $P_\infty$ exhibits a continuous second-order transition and the susceptibility $\chi$ peaks at criticality (see Fig.~\ref{fig:scaleInv}a). Finite-size scaling yields exponents consistent with $\beta=\gamma=1/3$ across different values of $\alpha$ (Fig.~\ref{fig:scaleInv}b; matching those of random monopartite networks, see SI1). The analytical threshold, derived from the bipartite percolation condition for Poisson random graphs \cite{Newman2001random}, is,
\[
k_c(\alpha)=\frac{2\sqrt{\alpha}}{1+\alpha},
\]
which agrees with numerical estimates (Fig.~\ref{fig:scaleInv}d). These results show that structural imbalance does not alter the universality class of the transition.

Despite this, the internal geometry of the critical component is not universal but depends on class imbalance. As $\alpha$ decreases, a secondary peak emerges in the susceptibility (Fig.~\ref{fig:scaleInv}c), revealing the onset of an additional structural scale associated with diffusion trapping at intermediate times, consistent with recent evidence that transport in bipartite systems is intrinsically constrained by their two-mode geometry~\cite{Jankowski2026}. This behavior is reminiscent of localization phenomena in diffusion on heterogeneous graphs, where spectral modes become spatially concentrated~\cite{Alt2024}, although here it arises from structural constraints rather than quenched disorder. At the same time, the organization of the critical backbone is systematically reshaped, even though the system remains locally tree-like. Hence, symmetric and strongly imbalanced systems share the same critical exponents while exhibiting distinct microscopic branching structures (i.e., fluctuations).

To test whether these differences persist under coarse-graining, we apply b-LRG at criticality (see SI1). All networks flow toward a tree-like fixed point, but the pathway depends on imbalance: heterogeneous, star-like structures in asymmetric systems are progressively integrated, whereas symmetric systems are already close to random trees. Imbalance thus controls how structural fluctuations are eliminated, without affecting the structural fixed point itself.

Critical bipartite networks thus offer a setting where universality and geometrical fluctuations can be cleanly separated. While large-scale scaling properties, linked to dimensionality, are fixed by the universality class, the geometry of the critical backbone—the structure governing diffusion and hierarchy—depends sensitively on two-mode constraints and partition imbalance.

\begin{figure*}
    \centering
    \includegraphics[width=0.95\textwidth]{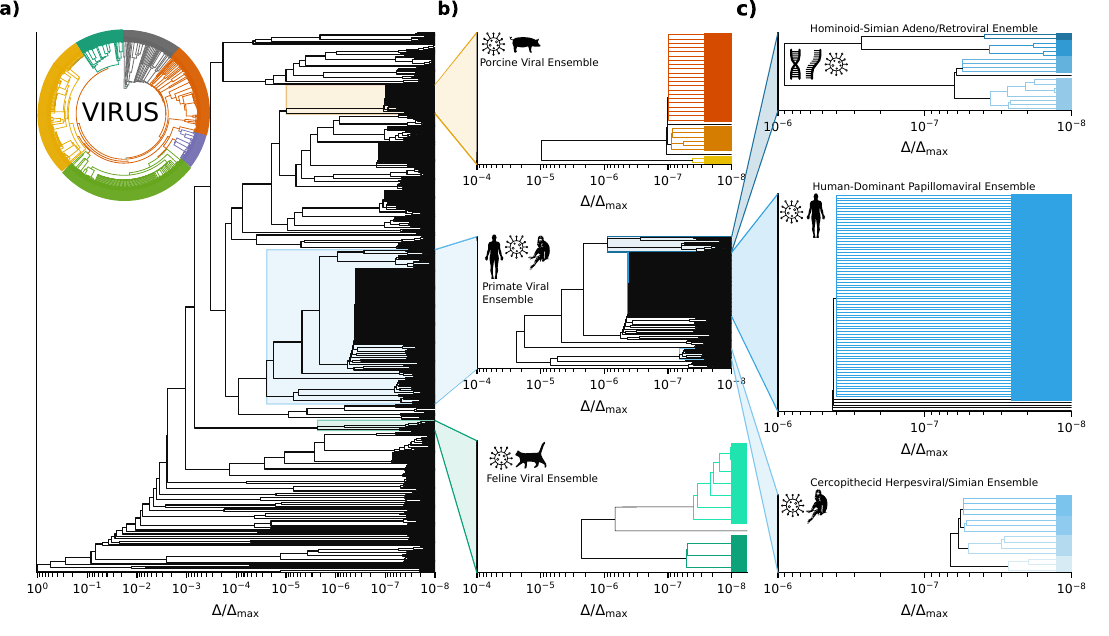}
    \caption{\textbf{Multiscale hierarchy of virus–host interactions.}
\textbf{(a)} Global dendrogram of viruses obtained from b-LRG, revealing a hierarchical organization across scales. The outer ring indicates viral taxonomic groups. Large-scale branches do not align with taxonomy, indicating that interaction-derived structure captures an organization distinct from phylogenetic similarity. 
\textbf{(b)} Intermediate-scale zooms of selected branches. Subtrees exhibit structured clustering associated with shared host environments and interaction patterns (icons), highlighting the emergence of ecological organization at mesoscopic scales. 
\textbf{(c)} High-resolution views of representative submodules. At small scales, the hierarchy becomes increasingly homogeneous, with dense and weakly differentiated clusters (colored blocks), reflecting convergence toward locally uniform interaction patterns. Branch length is expressed in normalized diffusion distance $\Delta/\Delta_{\max}$ (see Methods).}
    \label{fig:multiscale_dendrogram}
\end{figure*}

\subsection*{Multiscale hierarchy in bipartite networks and its distortion under projection}

Host--virus interaction networks are intrinsically bipartite: viral genomes interact with host species through cross-class associations, with no direct virus--virus or host--host links at the primary level \cite{Olival2017}. They thus provide a natural testbed to detect multiscale organization in a bipartite system and assess whether it is preserved under one-mode projection.

Applying b-LRG to the full bipartite structure yields a diffusion-based dendrogram (Fig.~\ref{fig:multiscale_dendrogram} for the virus class; see SI2 for the hosts class) characterized by a rich multiscale hierarchy. To assess its origin, we compare this interaction-derived hierarchy with independent taxonomic classifications (see Methods). For hosts, the inferred structure shows a weak but significant correlation with taxonomy (Spearman $r = 0.098$, Mantel test $p = 2.0 \times 10^{-4}$), indicating that evolutionary relatedness leaves only a limited imprint on interaction patterns. For viruses, no correlation is detected (Spearman $r = -0.023$, $p = 0.47$), showing that the hierarchy is statistically independent of viral taxonomy. The observed organization, therefore, reflects ecological positioning in the host--virus interaction space rather than phylogenetic similarity. 
Indeed, the inferred diffusion geometry partitions viruses into functionally cohesive clusters across scales. At large scales, it identifies broad host-associated branches, most prominently a primate-centered module that groups human pathogens (e.g., HIV, human papillomaviruses) with non-human primate counterparts (e.g., simian foamy viruses) and zoonotic agents (e.g., Ebolaviruses, SARS-related coronaviruses). At finer resolutions, the hierarchy isolates specialized sub-lineages, including tightly clustered swine-specific (e.g., African swine fever) and feline-specific (e.g., feline immunodeficiency and leukemia viruses) groups. This multiscale branching reflects ecological constraints encoded in host-sharing structure, rather than phylogenetic similarity.

By contrast, projecting the bipartite network onto a virus--virus graph—where two viruses are connected if they share hosts—and applying standard LRG produces a markedly different dendrogram (see SI2). Projection collapses the interaction space by reducing connectivity to shared-neighbor overlap, effectively collapsing alternating paths of length two into direct intra-class links \cite{Newman2001,Saracco2015}. Correlations mediated by longer paths are then retained only indirectly, and therefore with a distorted weight in the induced geometry. As a result, projection enhances dense clustering while obscuring the role-separated geometry of the original system. In contrast, our framework retains alternating paths of arbitrary length, allowing b-LRG to reconstruct distances from the full interaction space.
This difference persists across scales. At large scales (Fig.~\ref{fig:multiscale_dendrogram}a), bipartite analysis reveals structural divisions that are not recovered under projection. At intermediate and small scales (Fig.~\ref{fig:multiscale_dendrogram}b,c), projection compresses the branching structure, whereas bipartite renormalization preserves heterogeneous sub-lineages and scale-dependent differentiation.

\begin{figure*}[]
    \centering
    \includegraphics[width=1.8\columnwidth]{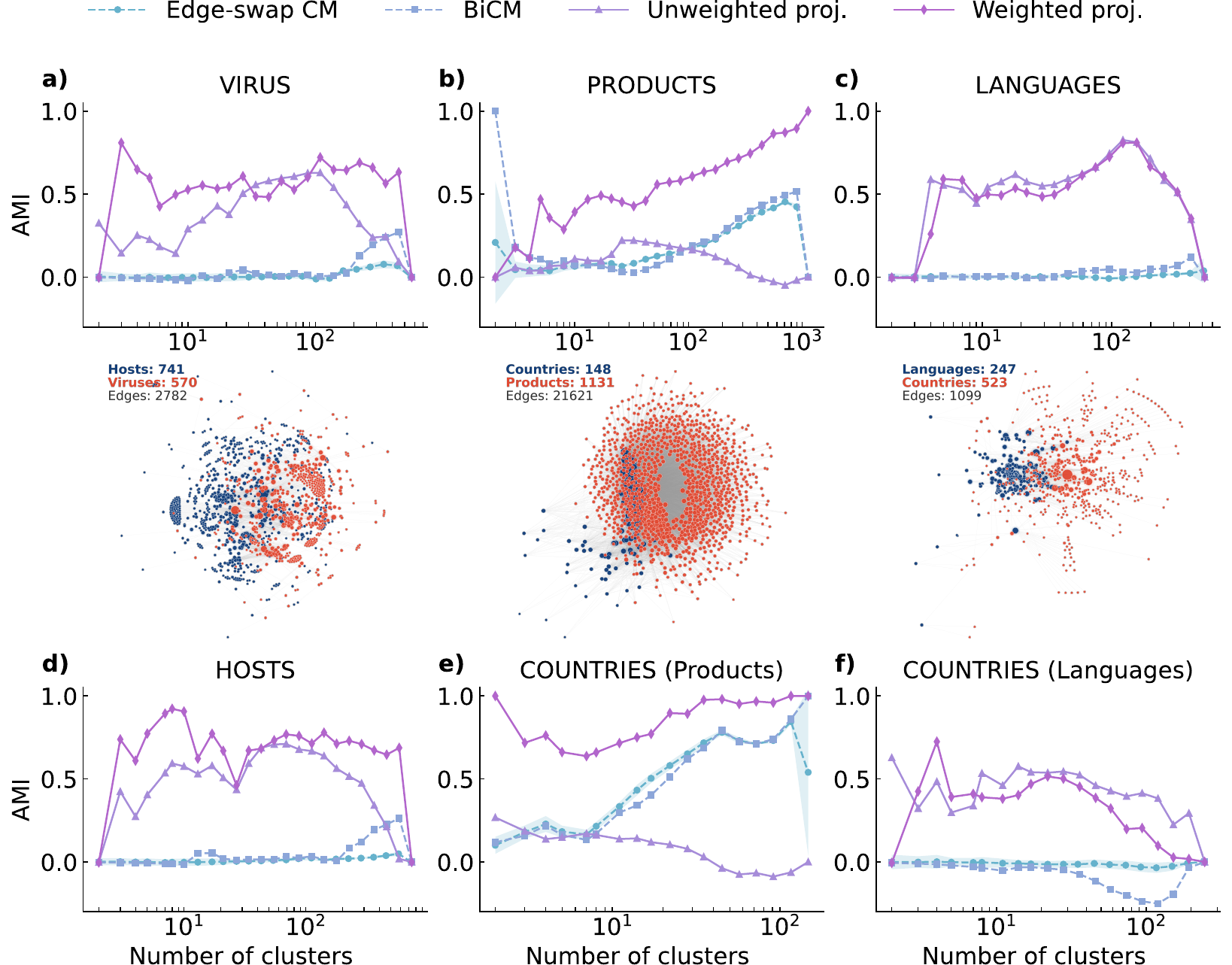}
    \caption{\textbf{One-mode projection distorts multiscale organization.}
Adjusted mutual information (AMI) between clusters obtained from the empirical bipartite dendrogram constructed by b-LRG and those derived from degree-preserving bipartite null models (edge-swap CM and BiCM, again constructed by b-LRG) or one-mode projections (unweighted and weighted, constructed by standard LRG), as a function of the number of clusters.
We analyze three empirical bipartite systems: host--virus interactions \cite{Olival2017}, country--product export data \cite{Hidalgo2007,Tacchella2012,Caccioli2014}, and country--language associations \cite{Kunegis2013}.
\textbf{(a--c)} Virus, product, and language partitions.
\textbf{(d--f)} Host, country-product, and country-language partitions.
Across all systems and partitions, degree-preserving null models yield AMI values close to zero over most resolutions, indicating that the empirical hierarchy cannot be explained by degree constraints alone.
One-mode projections retain partial similarity but fail to recover the full multiscale geometry of the bipartite interaction space.}
    \label{fig:NullModel}
\end{figure*}

We quantify these differences by comparing clusters obtained from the empirical bipartite dendrogram with those derived from one-mode projections, using adjusted mutual information (AMI) across dendrogram cuts (Fig.~\ref{fig:NullModel}a,d). 
As a benchmark, we also consider dendrograms induced by bipartite degree-preserving null models (edge-swap CM \cite{Strona2014} and BiCM \cite{Saracco2015}), commonly used to filter spurious connections in projected networks \cite{Tumminello2011,Gualdi2016,Saracco2017,Tamarit2020,Cimini2022,Neal2014,Neal2024}. 
The AMI between partitions of empirical and null bipartite networks collapses to values close to zero across nearly all resolutions, indicating that the observed hierarchy cannot be explained by degree constraints alone. 
By contrast, the AMI between partitions of the original bipartite and the projected networks remains positive over broad ranges of resolutions, but rarely reaches values close to 1. 
Projection is therefore not equivalent to the bipartite representation, nor completely uninformative, by reproducing the full bipartite hierarchy only partially. 
This observation is confirmed across several empirical networks \cite{Barrett1987,Bartomeus2008,Schleuning2011} (Fig.~\ref{fig:NullModel}b,c,e,f and SI3). The same comparison using normalized mutual information, which measures raw partition similarity without correcting for chance, leads to consistent conclusions (see SI3).

The effectiveness of the projection in recovering the bipartite hierarchy depends on how much of the organization is captured by local overlap patterns. When dominant correlations are encoded in short paths, one-mode projections retain the information of the full bipartite hierarchy, as in the Davis Southern Women network (see SI3).
By contrast, in larger, heterogeneous and hierarchically organized networks, projection fails once multi-step correlations become relevant. In bipartite systems, these correspond to even-length alternating paths, 
which encode dependencies beyond pairwise overlap. Collapsing the two-mode structure removes these pathways, distorting the diffusion geometry and, consequently, the inferred hierarchy.

\section{Conclusion}
Multiscale organization in bipartite systems is not, in general, recoverable from projected representations. By introducing effective intra-class links, projection modifies diffusion pathways and therefore the geometry from which hierarchy emerges \cite{Newman2001,Saracco2015}. As a result, coarse-graining performed on projected networks does not preserve the multiscale structure of the original system.
This reflects a more general mechanism: projection effectively truncates correlations carried by alternating interaction paths beyond nearest-neighbor overlap \cite{Newman2001}. In this sense, projection acts as a low-order truncation of the full diffusion geometry, retaining only the shortest alternating paths. In bipartite systems, these paths encode dependencies across multiple scales; removing one partition eliminates these contributions and reshapes the effective distances that define hierarchical organization \cite{Jankowski2026}.

Treating bipartiteness as a deviation from a monopartite structure leads to systematic distortions, replacing the full cross-class organization with a reduced and biased representation. More broadly, our results show that in role-differentiated systems, scale is not a property of either partition in isolation, but of the full interaction architecture. Removing one partition does not simply reduce the system, but alters the structure from which organization emerges. This reflects the fact that multiscale organization is inherently a joint property of the full system. In this sense, one may view the two-mode architecture as irreducible: “a part torn out from the whole cannot be made the whole, nor can it be understood apart from the whole” \cite{Bogdanov1996}.

The analysis of host--virus interactions, together with comparisons across multiple empirical systems, shows that these effects are not merely formal. Projected representations compress branching structure and obscure cross-class organization, whereas bipartite renormalization reveals an interaction-derived hierarchy that is only weakly related to taxonomy and therefore captures ecological rather than phylogenetic organization.

These findings point to a general principle: multiscale organization is not invariant under representation when representation alters the geometry through which interactions propagate. In systems with structured interaction constraints—such as role-differentiated, layered, or typed architectures—coarse-graining must preserve these constraints to retain the geometry that defines scale. Projection is therefore not merely a reduction of complexity, but a transformation that truncates the underlying diffusion geometry, selectively retaining short-range correlations while suppressing higher-order structure. As a result, the organization that emerges across scales can be systematically reshaped. This highlights a broader limitation of representational reductions in complex systems, suggesting that preserving interaction structure is essential whenever coarse-graining is used to infer multiscale behavior.

The framework introduced here provides a principled approach to study scale-dependent organization, stability, and information flow in two-mode systems \cite{LRG,InfoCore} without relying on geometric embeddings or null-model assumptions. Future work may extend this framework to dynamical processes, robustness, and critical phenomena in role-differentiated architectures, as well as to multilayer systems with multiple structural constraints. 


\emph{\textbf{Acknowledgments--}}We thank A.Gabrielli and F.Saracco for useful discussions and comments. 
G. C. acknowledges financial support from the Italian Ministry of University and Research (MUR) through the National Recovery and Resilience Plan (NRRP) - NextGenerationEU: 
projects C2T (code P2022E93B8, CUP E53D23018320001) and RENet (code 2022MTBB22, CUP E53D23001770006). 
P.V. acknowledges the Spanish Ministry and Agencia Estatal de Investigaci\'on (AEI), MICIN/AEI/10.13039/501100011033, for financial support through Project PID2023-149174NB-I00, funded also by ERDF/EU.\vspace{-2mm}

\section{Methods}

\paragraph{\textbf{Bipartite Laplacian renormalization.}}

Let $G = (V_A \cup V_B, \cal{E})$ be a bipartite network with partitions $V_A$ and $V_B$, number of nodes in each class $|V_A| = N_A$, $|V_B| = N_B$, and number of links $|\cal{E}|=E$. The $(N_A+N_B)\times(N_A+N_B)$ adjacency matrix $\hat A$ encodes connections among nodes. The graph Laplacian is $\hat L = \hat D - \hat A$, where $\hat D$ is the diagonal degree matrix. Diffusion dynamics on the network follow $\dot{\mathbf{s}}(\tau) = - \hat L \mathbf{s}(\tau)$, with propagator $\hat K(\tau) = e^{-\tau \hat L}$. The Laplacian density matrix
\[
\hat \rho(\tau) = \frac{e^{-\tau \hat L}}{\mathrm{Tr}\, e^{-\tau \hat L}},
\]
provides a trace-normalized measure of diffusion-based communication across the network.
An effective distance between nodes is defined as
\[
\mathcal{D}_{ij}(\tau) = \frac{1 - \delta_{ij}}{\rho_{ij}(\tau)},
\]
so that nodes with stronger diffusion-mediated communication are closer. This distance induces a hierarchical geometry on the network.

To preserve the bipartite constraint under coarse-graining, we impose an infinite geometric penalty on inter-class pairs, defining
\[
\mathcal{D}^{B}_{ij}(\tau)
=
\frac{1 - \delta_{ij}}{\rho_{ij}(\tau)} \, \mathbf{1}_{\mathrm{intra\text{-}class}}
+
\frac{1}{\varepsilon} \, \mathbf{1}_{\mathrm{inter\text{-}class}},
\qquad
\varepsilon \to 0^{+}.
\]
In this limit, clustering is restricted to nodes within the same partition, preserving role separation at all scales.

\paragraph{\textbf{Dendrogram and Spectral gap.}}

The distance matrix $\mathcal{D}^{B}(\tau)$ is used as input for hierarchical clustering (average linkage hierarchical clustering was used throughout).
To quantify the statistical significance of hierarchical branching, we evaluate the gap function $\Psi(n)$, defined as the relative separation between consecutive linkage distances in the dendrogram. Specifically,
\[
\Psi(n) = \frac{\Delta_{n+1} - \Delta_n}{\Delta_n},
\]
where $\Delta_n$ denotes the linkage distance at the $n$-th merging step. Peaks in $\Psi(n)$ identify robust partitions, corresponding to robust branching levels in the hierarchy.

\paragraph{\textbf{Spectral specific heat.}}

The spectral specific heat is computed from the Laplacian density matrix as
\[
C(\tau) = -\tau \frac{d}{d\tau} \mathrm{Tr}\left[ \hat \rho(\tau) \log \hat \rho(\tau) \right],
\]
where $\mathrm{Tr}\!\left[ \hat \rho(\tau) \log \hat \rho(\tau) \right]$ is the spectral entropy. This quantity captures the redistribution of diffusion modes across scales and is used to identify scale-invariant regimes and mesoscale heterogeneity. 

\paragraph{\textbf{Random bipartite network model.}}

We generate bipartite random graphs with fixed partition sizes $N_A$, $N_B$ and number of links $E$, corresponding to a microcanonical ensemble. The asymmetry parameter is defined as $\alpha = N_A / N_B$. For fixed $\alpha$ and system size $N = N_A + N_B$, the average node degree reads
\[
\langle k \rangle = \frac{2E}{N}.
\]
Results were averaged over an ensemble of $10^3$ random networks.

\paragraph{\textbf{Percolation observables.}}

The order parameter is defined as the fraction of nodes in the largest connected component,
\[
P_\infty = \frac{S_{\mathrm{max}}}{N}.
\]
The susceptibility is given by
\[
\chi = N \left( \langle P_\infty^2 \rangle - \langle P_\infty \rangle^2 \right),
\]
where averages are taken over the ensemble.

The critical point $k_c$ was estimated from the location of the maximum of $\chi$ for each system size, and compared to the  analytical threshold
\[
k_c(\alpha) = \frac{2\sqrt{\alpha}}{1+\alpha}.
\]

\paragraph{\textbf{One-mode projections.}}

Monopartite projections of a bipartite network connect two nodes of the same class if they share at least one common neighbor in the other class. Edge weights correspond to the number of shared neighbors; binarized projections retain only the presence or absence of links. Projected networks are analyzed using the same Laplacian-based diffusion distance and hierarchical clustering procedure as in the bipartite case \cite{Modularity}.

\paragraph{\textbf{Null models.}}

Degree-preserving bipartite null models were generated using (i) the bipartite configuration model (BiCM), which preserves the degree sequences of both partitions on average \cite{Saracco2015}, and (ii) an edge-swap randomization procedure that preserves the exact bipartite degree sequences \cite{Strona2014}. These models serve as statistical benchmarks to evaluate how much of the inferred hierarchical structure can be accounted for solely by degree constraints.

\paragraph{\textbf{Information-theoretic comparison of dendrograms.}}

To compare hierarchies across representations, we cut each dendrogram at a fixed number of clusters and compared the resulting partitions $\mathcal{U}$ and $\mathcal{V}$. Their normalized mutual information is
\[
\mathrm{NMI}(\mathcal{U},\mathcal{V})
=
\frac{2 I(\mathcal{U},\mathcal{V})}
{H(\mathcal{U})+H(\mathcal{V})},
\]
where $I$ is the mutual information and $H$ the partition entropy. Thus, $\mathrm{NMI}=1$ for identical partitions and approaches zero for independent ones. Since NMI does not correct for chance agreement, we also compute
\[
\mathrm{AMI}(\mathcal{U},\mathcal{V})
=
\frac{
I(\mathcal{U},\mathcal{V})-\mathbb{E}[I]
}{
\frac{1}{2}\left[H(\mathcal{U})+H(\mathcal{V})\right]-\mathbb{E}[I]
},
\]
where $\mathbb{E}[I]$ is the expected mutual information for random partitions with the same cluster sizes. We report AMI in the main text as a chance-corrected measure of whether projected or randomized representations recover the empirical bipartite hierarchy.

\paragraph{\textbf{Taxonomic comparison.}}

Independent host and viral taxonomic hierarchies were obtained from standard classification databases \cite{Olival2017}. Cophenetic distance matrices derived from interaction-based dendrograms were compared with taxonomic distance matrices using Spearman rank correlation. Statistical significance was assessed via Mantel tests (Spearman correlation) with $5{,}000$ permutations.

%

\clearpage


\section*{Supplementary material}

# Two-mode geometry controls multiscale organization in bipartite systems
# Supplementary Materials

Ottavia Falconi,[1,2] Giulio Cimini,[3,2] and Pablo Villegas[2,4,*]

[1] *Physics Department University of Rome Tor Vergata, 00133 Rome, Italy*
[2] *Enrico Fermi Research Center (CREF), 00184, Rome, Italy*
[3] *Physics Department and INFN, University of Rome Tor Vergata, 00133 Rome, Italy*
[4] *Instituto Carlos I de Física Teórica y Computacional, Universidad de Granada, 18071 Granada, Spain*

## CONTENTS



* pablo.villegas@cref.it

# 1. PERCOLATION TRANSITION AND FINITE-SIZE SCALING IN IMBALANCED BIPARTITE NETWORKS

## 1.1. Random bipartite network generation

To probe how partition asymmetry affects the internal geometry of the critical backbone, we study random bipartite networks composed of two disjoint sets of nodes, $A$ and $B$, with sizes $N_A$ and $N_B$. The total number of nodes is $N = N_A + N_B$, and the partition asymmetry is defined as $\alpha = N_A/N_B \leq 1$.

The network ensembles are generated using the *igraph* library by fixing the number of nodes in each set and the total number of links $L$, corresponding to a microcanonical bipartite ensemble. The total number of links is related to the mean connectivity (average degree) $\langle k \rangle$ through

$$L = \frac{\langle k \rangle N}{2}\,, \tag{1}$$

where the factor of 1/2 accounts for the fact that each link contributes to the degree of two nodes. Using $\langle k \rangle$ as the control parameter allows direct comparison across different system sizes and partition asymmetries, and enables precise tuning to criticality.

## 1.2. Analytical percolation threshold via generating functions

We derive the percolation threshold of the bipartite ensemble using the generating function formalism. Let $f_0(x)$ and $g_0(x)$ denote the probability generating functions of the degree distributions of nodes in sets $A$ and $B$, respectively. The corresponding mean degrees are $k_A = f_0'(1)$ and $k_B = g_0'(1)$.

Following a randomly chosen edge, the excess degree distributions are generated by $f_1(x) = f_0'(x)/k_A$ and $g_1(x) = g_0'(x)/k_B$. In bipartite networks, the generating function for the number of second neighbors is given by the composition $f_0(g_1(x))$. The mean component size diverges when the expected number of second neighbors reached from a given edge exceeds unity, yielding the condition

$$f_1'(1)g_1'(1) = 1 \quad \Rightarrow \quad \frac{f_0''(1)}{f_0'(1)}\frac{g_0''(1)}{g_0'(1)} = 1\,. \tag{2}$$

In the random bipartite ensemble considered here, the degree distributions converge to Poisson distributions in the thermodynamic limit ($N \to \infty$), with $f_0(x) = e^{k_A(x-1)}$ and $g_0(x) = e^{k_B(x-1)}$. Substituting into the condition above yields the criticality criterion

$$k_A k_B = 1\,. \tag{3}$$

Expressing the mean degrees in terms of the global mean connectivity $\langle k \rangle = 2L/(N_A + N_B)$, we have $k_A = \frac{\langle k \rangle (N_A+N_B)}{2N_A}$ and $k_B = \frac{\langle k \rangle (N_A+N_B)}{2N_B}$. Introducing the asymmetry parameter $\alpha = N_A/N_B$, we obtain

$$\langle k \rangle_c(\alpha) = \frac{2\sqrt{\alpha}}{1+\alpha}\,. \tag{4}$$

This expression allows direct tuning of systems with different imbalance $\alpha$ to their respective critical points, enabling controlled comparisons of their internal geometry at fixed criticality.

### 1.3. Finite-size scaling and universality

To assess whether partition imbalance affects the universality class of the transition, we perform a finite-size scaling (FSS) analysis. We measure the fraction of nodes in the largest connected component, $P_\infty(N, \langle k \rangle)$, and the susceptibility, $\chi(N, \langle k \rangle)$.

Standard FSS theory predicts that these observables follow the scaling forms

$$P_\infty(N, \langle k \rangle) = N^{-\beta/\bar{\nu}} \mathcal{F}_P \left( (\langle k \rangle - \langle k \rangle_c(\alpha)) N^{1/\bar{\nu}} \right), \tag{5}$$

$$\chi(N, \langle k \rangle) = N^{\gamma/\bar{\nu}} \mathcal{F}_\chi \left( (\langle k \rangle - \langle k \rangle_c(\alpha)) N^{1/\bar{\nu}} \right), \tag{6}$$

where $\gamma$, $\beta$, and $\bar{\nu}$ are the corresponding finite-size scaling exponents.

At criticality ($\langle k \rangle = \langle k \rangle_c$), these expressions reduce to power-law scaling with system size:

$$P_\infty(N, \langle k \rangle_c) \propto N^{-\beta/\bar{\nu}}, \quad \chi_{\max}(N) \propto N^{\gamma/\bar{\nu}}. \tag{7}$$

Supplementary Figs. 1 and 2 show the scaling behavior for two imbalanced configurations, $\alpha = 0.25$ and $\alpha = 0.11$. As system size increases, the transition sharpens and the susceptibility peak grows, consistent with a continuous phase transition.

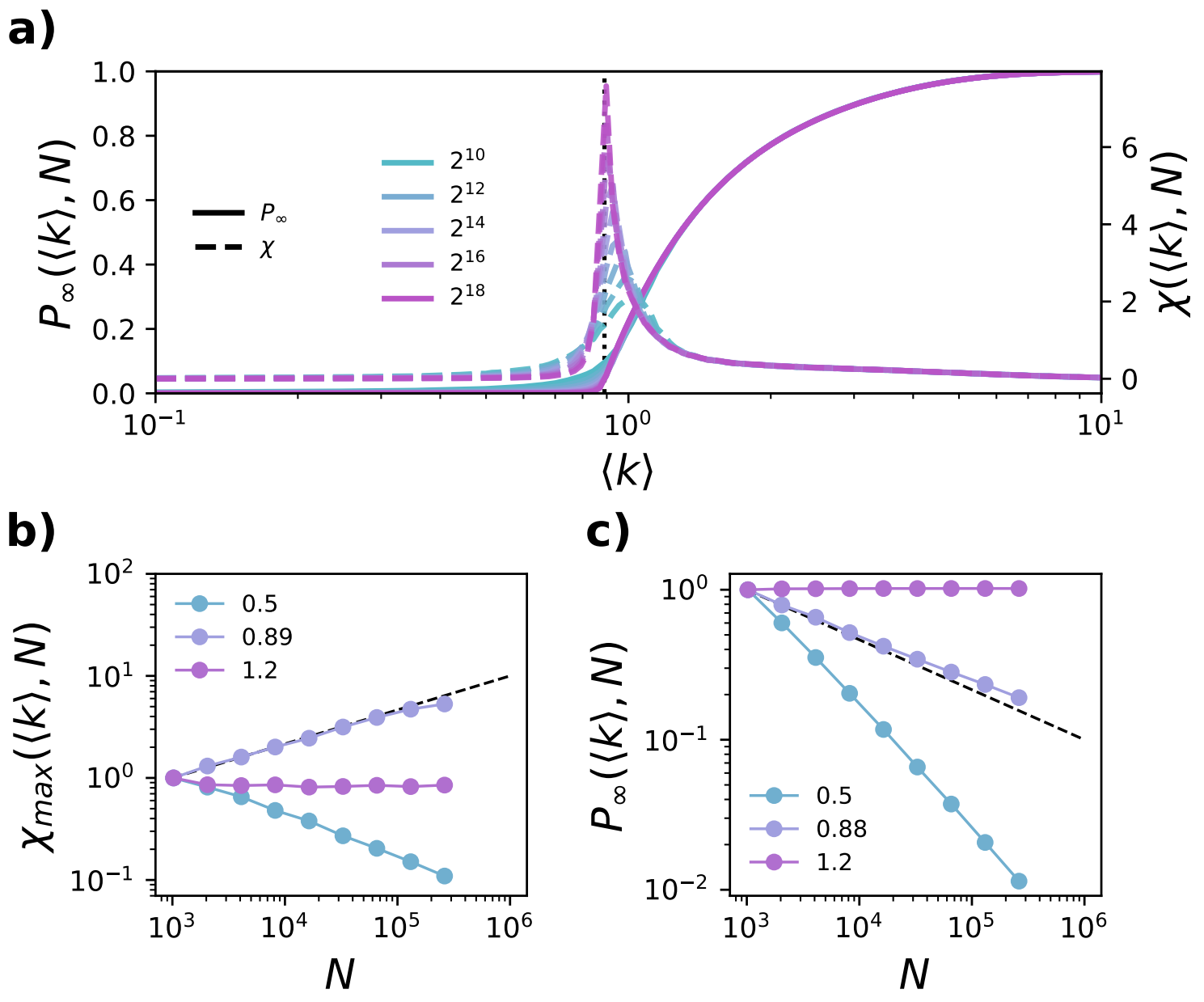


Figure 1. **Finite-size scaling analysis of Random Bipartite Networks with $\alpha = 0.25$. a)** Fraction of nodes in the giant component $P_\infty$ (solid lines) and susceptibility $\chi$ (dashed lines) as a function of the mean connectivity $\langle k \rangle$ for different system sizes $N$ ($2^{10}$–$2^{18}$). The vertical dotted line marks the analytical percolation threshold $\langle k \rangle_c$. As $N$ increases, the transition sharpens, and the susceptibility peak grows. **b)** Maximum susceptibility $\chi_{\max}$ as a function of system size $N$ (log–log scale) for subcritical, critical, and supercritical regimes. At criticality, $\chi_{\max} \propto N^{\gamma/\bar{\nu}}$. The dashed line indicates the mean-field scaling $\gamma/\bar{\nu} = 1/3$. **c)** System-size dependence of the order parameter $P_\infty$ at selected values of $\langle k \rangle$. At criticality, $P_\infty \propto N^{-\beta/\bar{\nu}}$. The dashed line indicates the mean-field scaling $-\beta/\bar{\nu} = -1/3$. The consistency of the scaling in (b) and (c) indicates that partition imbalance does not affect the universality class.

From the scaling of $\chi_{\max}$ and $P_\infty$ at criticality, we extract exponents $\gamma/\bar{\nu} = 1/3$ and $\beta/\bar{\nu} = 1/3$ for both values of $\alpha$, in agreement with mean-field predictions. These results indicate that partition imbalance does not modify the universality class of the transition.

This invariance shows that the additional structural features observed at strong imbalance—such as secondary susceptibility peaks and heterogeneous branching patterns—cannot be attributed to a change in critical behavior, but to modifications of the underlying network geometry. In this sense, imbalance affects the multiscale structure of the critical backbone without altering its macroscopic scaling properties.

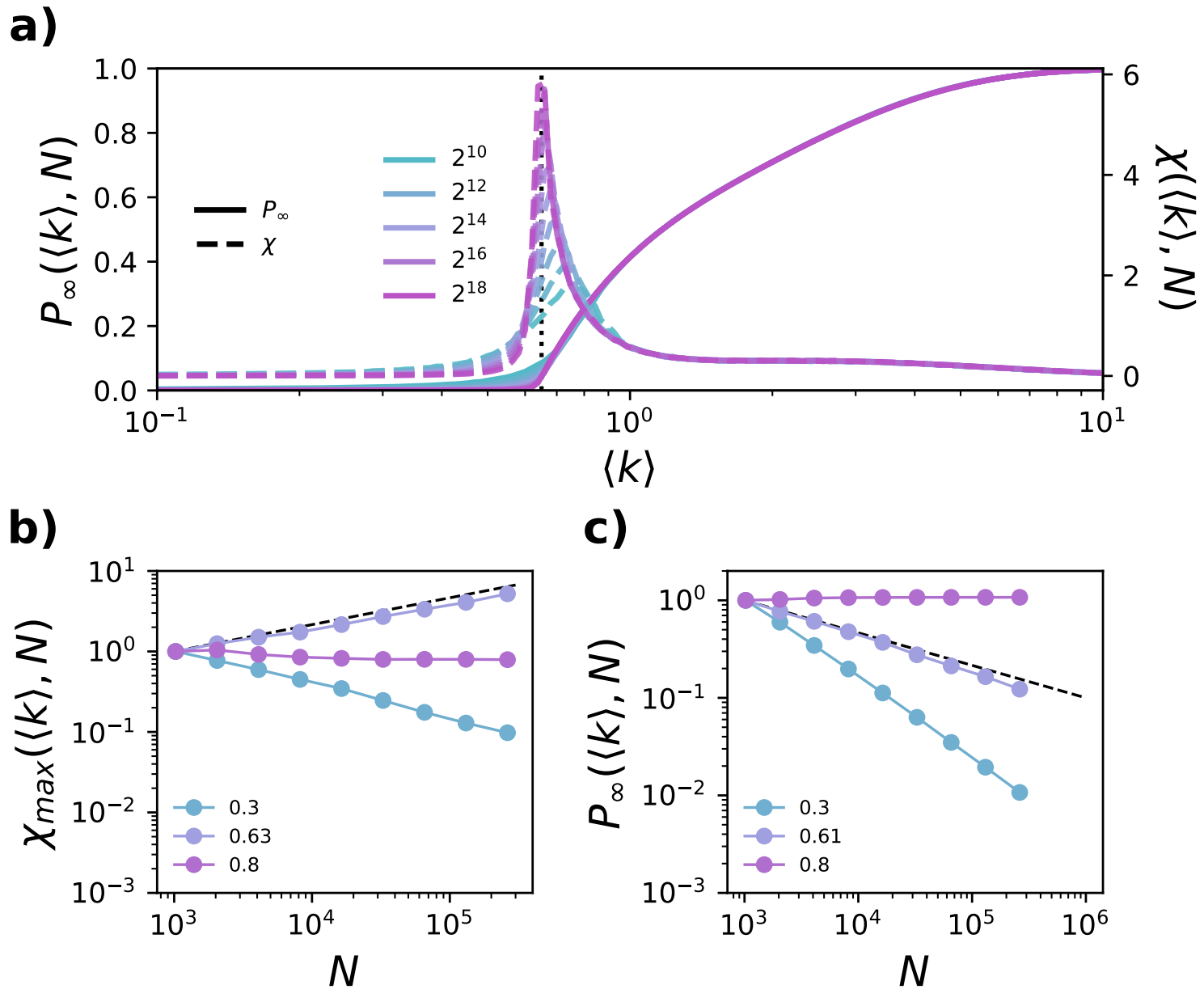


Figure 2. **Finite-size scaling analysis of Random Bipartite Networks with** $\alpha = 0.11$**. a)** Fraction of nodes in the giant component $P_\infty$ (solid lines) and susceptibility $\chi$ (dashed lines) as a function of the mean connectivity $\langle k \rangle$ for different system sizes $N$ ($2^{10}$–$2^{18}$). The vertical dotted line marks the analytical percolation threshold $\langle k \rangle_c$. As $N$ increases, the transition sharpens, and the susceptibility peak grows. **b)** Maximum susceptibility $\chi_{\max}$ as a function of system size $N$ (log–log scale) for different values of $\langle k \rangle$ (below, at, and above criticality). At criticality, $\chi_{\max} \propto N^{\gamma/\bar{\nu}}$. The dashed line indicates the mean-field scaling $\gamma/\bar{\nu} = 1/3$. **c)** System-size dependence of the order parameter $P_\infty$ at selected values of $\langle k \rangle$. At criticality, $P_\infty \propto N^{-\beta/\bar{\nu}}$. The dashed line indicates the mean-field scaling $-\beta/\bar{\nu} = -1/3$. The consistency of the scaling in (b) and (c) indicates that partition imbalance does not affect the universality class.

### 1.4. b-LRG flow and multiscale geometry

To probe how these differences evolve across scales, we apply the bipartite LRG transformation at criticality for different values of $\alpha$. While the finite-size scaling analysis shows that partition imbalance does not affect the universality class of the transition, the internal multiscale geometry of the critical backbone depends sensitively on $\alpha$. To probe this, we apply the b-LRG framework to the bipartite ensembles, which resolves structure across scales through the diffusion time $\tau$.

The spectral specific heat $C(\tau)$ (entropy susceptibility) characterizes how diffusion modes are distributed across scales. A plateau in $C(\tau)$ indicates scale-invariant diffusion over a range of $\tau$. As shown in Supplementary Fig. 3(a), the symmetric case ($\alpha = 1$) exhibits such a plateau, consistent with the one expected for critical random trees. In this regime, the backbone shows a homogeneous, tree-like branching structure across scales.

By contrast, under strong imbalance ($\alpha \ll 1$), the distribution of diffusion modes shifts toward smaller scales. In particular, in highly imbalanced systems (e.g., $\alpha = 0.025$, Supplementary Fig. 3b), $C(\tau)$ develops a pronounced peak at low $\tau$, signaling the emergence of a characteristic microscopic scale. This scale is associated with heterogeneous, hub-like structures within the larger partition, which locally slow down diffusion and introduce an additional structural level. This scale also identifies the resolution at which these structures are integrated under coarse-graining. As illustrated in Supplementary Fig. 3(b), grouping nodes within these local structures at the corresponding $\tau$ produces an effective coarse-grained network in which these heterogeneities are absorbed into larger units.

Despite these differences at small scales, the renormalization flow leads to a common mesoscopic structure: both symmetric and imbalanced systems approach a tree-like backbone under successive coarse-graining. However, the pathway to this structure depends on imbalance, reflecting differences in how local heterogeneities are organized and integrated across scales.

These results show that partition imbalance reshapes the multiscale geometry of the critical network without modifying its large-scale scaling behavior, reinforcing the separation between universality and geometrical fluctuations discussed in the main text.

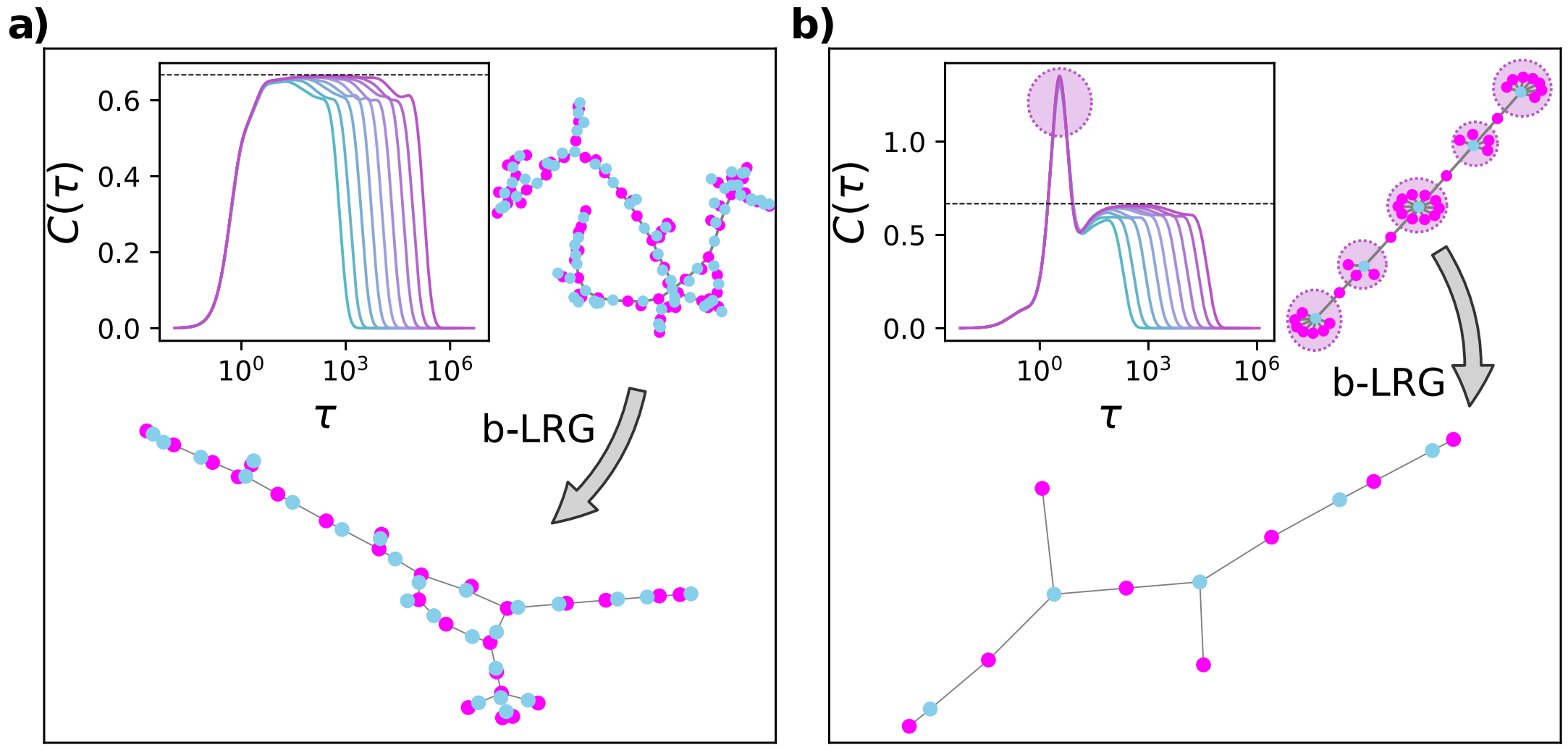


Figure 3. **Multiscale geometry and Bipartite Laplacian Renormalization Group (b-LRG) flow of critical bipartite networks. a)** Symmetric bipartite network ($\alpha = 1$). The inset shows the spectral specific heat $C(\tau)$ for different system sizes, exhibiting a broad plateau consistent with scale-invariant diffusion. The backbone displays a homogeneous, tree-like structure that is preserved under coarse-graining (grey arrow). **b)** Strongly imbalanced network ($\alpha = 0.025$). The inset shows that $C(\tau)$ develops a peak at low $\tau$, indicating a characteristic microscopic scale associated with heterogeneous, hub-like structures (highlighted). Coarse-graining at this scale integrates these structures, yielding a more homogeneous mesoscopic backbone.

## 2. HOST MULTISCALE HIERARCHY AND TAXONOMIC CORRELATION

In the main text, we showed that bipartite diffusion geometry reveals a multiscale hierarchy for viral species that is distorted under one-mode projection. Here, we extend the analysis to the host partition within the same interaction space.

Unlike projected representations, which compress the branching structure into dense modules driven by shared-virus overlap, the bipartite Laplacian Renormalization Group (b-LRG) preserves cross-class pathways. The resulting host dendrogram (shown in Supplementary Fig. 4) exhibits heterogeneous sub-lineages and clear scale-dependent differentiation, reflecting the ecological positioning of hosts based on their viral associations.

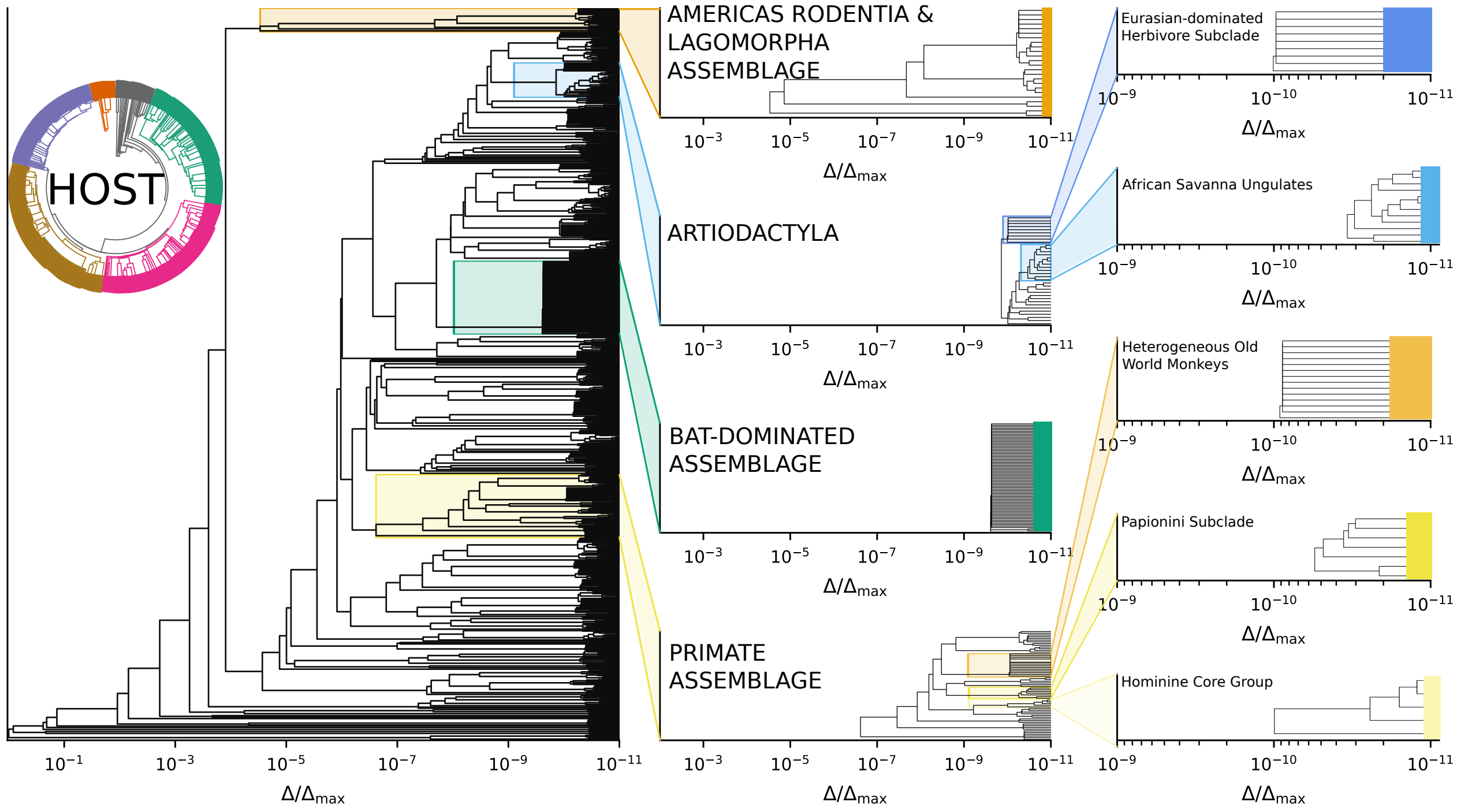


Figure 4. **Multiscale hierarchical organization of host species derived from bipartite diffusion geometry.** The dendrogram is obtained from the b-LRG applied to the host–virus bipartite network. The horizontal axis represents the normalized diffusion distance $\Delta/\Delta_{\max}$. At large scales, the hierarchy separates major host groups into broad clusters. Four major structural blocks emerge at large scales, corresponding to an Americas Rodentia and Lagomorpha assemblage (orange), an Artiodactyla assemblage (blue), a Bat-dominated assemblage (green), and a Primate assemblage (yellow). At finer scales, additional substructure emerges, revealing differentiation within these groups. Zoom-in panels reveal scale-dependent differentiation within these macro-groups. In particular, the Artiodactyla cluster separates Eurasian-dominated herbivores from African savanna ungulates, while the Primate assemblage distinguishes the Hominine core group from Papionini and other Old World monkeys, illustrating how interaction-based geometry resolves substructure beyond coarse groupings. The organization reflects interaction patterns rather than taxonomic classification, capturing ecological structure encoded in the bipartite interaction space.

At large scales, the hierarchy separates hosts into distinct interaction assemblages, including a Primate group, an Artiodactyla group, a bat-dominated assemblage (which also integrates sympatric carnivores and rodents), and a New World Rodentia and Lagomorpha group.

At finer resolutions, additional differentiation emerges within these macro-groups. For instance, the Artiodactyla cluster separates African savanna ungulates from Eurasian-dominated herbivores, while the Primate assemblage splits into distinct subclades, distinguishing the Hominine group from Papionini and other Old World monkeys. These patterns illustrate how interaction-based geometry resolves hierarchical substructure beyond coarse groupings.

To quantify this, we compare the interaction-derived dendrogram with independent phylogenetic classifications. While the viral hierarchy shows no statistical dependence on taxonomy (Spearman $r = -0.023$, $p = 0.47$), the host hierarchy exhibits a weak but statistically significant correlation (Spearman $r = 0.098$, Mantel test $p = 2.0 \times 10^{-4}$). This indicates that phylogeny leaves a detectable but limited imprint on the interaction structure, and confirms that the hierarchy is shaped by interaction patterns rather than purely by taxonomy.

Overall, the inferred hierarchy captures an ecological organization that is only partially aligned with evolutionary relatedness. This reflects the fact that host similarity in the interaction space is shaped not only by phylogeny but also by ecological and functional constraints.

Projection obscures this structure. As shown in Supplementary Fig. 5, both unweighted and weighted projections compress the branching hierarchy into dense structures, replacing diffusion-based distances with overlap-based similarity. This transformation removes cross-class pathways and suppresses the multiscale organization present in the bipartite representation.

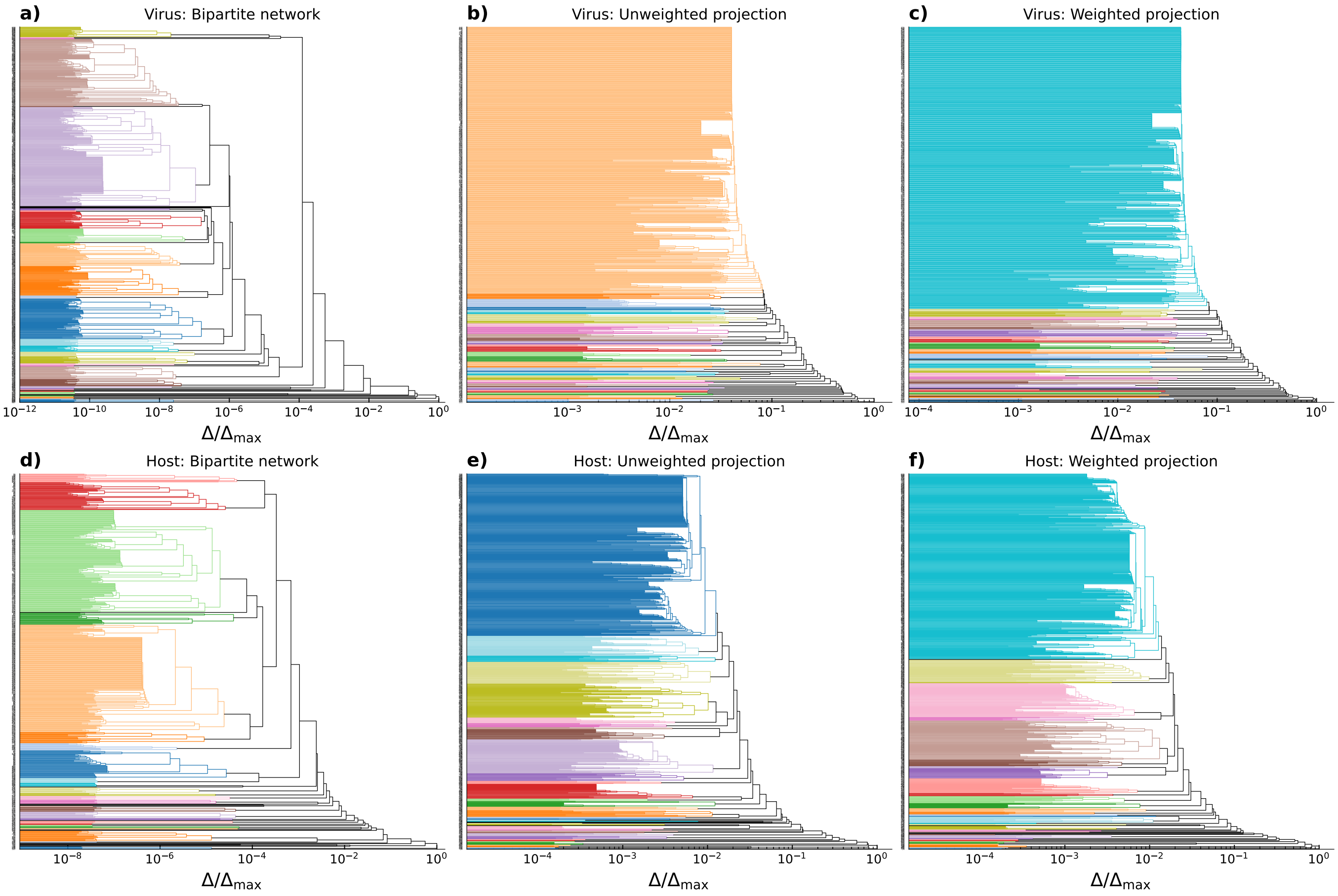


Figure 5. **Distortion of multiscale hierarchy under one-mode projection.** **a–c,** Viral dendrograms obtained from: (**a**) bipartite geometry, and from (**b**) unweighted (**c**) weighted projections. **d–f,** Corresponding host dendrograms. Bipartite representations preserve multiscale branching, whereas projections compress the hierarchy into dense structures, reflecting the loss of higher-order interaction pathways.

## 3. TOPOLOGICAL LIMITS AND STRUCTURAL DEPENDENCIES OF ONE-MODE PROJECTIONS

To investigate the conditions under which one-mode projections represent a well-behaved approximation, we extend the NMI and AMI analysis to a broader set of empirical bipartite networks. As discussed in the main text, projection acts as a low-order truncation of the diffusion geometry: it collapses alternating paths of length two into direct intra-class links while suppressing correlations encoded in longer alternating paths ($A-B-A-B$).

The impact of this truncation depends on the structural origin of multiscale organization. When the hierarchy is dominated by local overlap (short-range correlations), projections can approximate the bipartite geometry. When longer-range correlations are relevant, projection leads to systematic distortions.

### 3.1. Davis Southern Women: a low-complexity regime

We first consider the Davis Southern Women network, a benchmark bipartite system with relatively dense connectivity and limited hierarchical depth.

As shown in Supplementary Fig. 6, the network is dominated by two dense clusters with strong internal overlap.

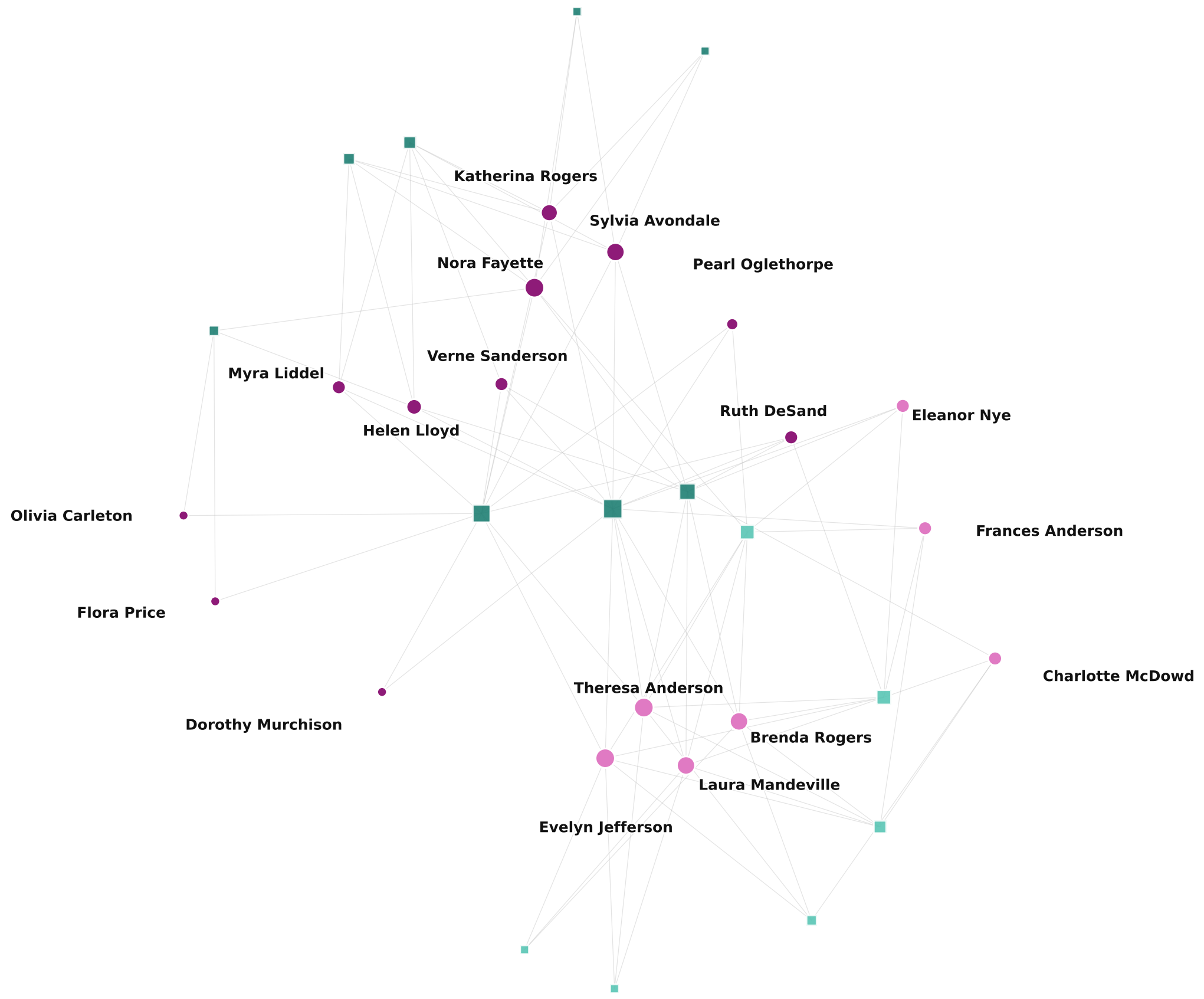


Figure 6. **Davis Southern Women network.** Bipartite network of women and events. Node colors indicate the four-group partition obtained from the diffusion-based hierarchical analysis.

In this regime, the diffusion geometry is largely determined by shared-neighbor overlap. Because correlations are primarily encoded in short alternating paths, projection does not substantially alter the effective geometry.

Consistently, the weighted one-mode projection retains a high similarity with the bipartite hierarchy across resolutions (see Supplementary Fig. 7). This identifies a limiting regime in which weighted projection provides a good approximation of the bipartite geometry, namely when multiscale organization is governed by local density rather than multi-step correlations.

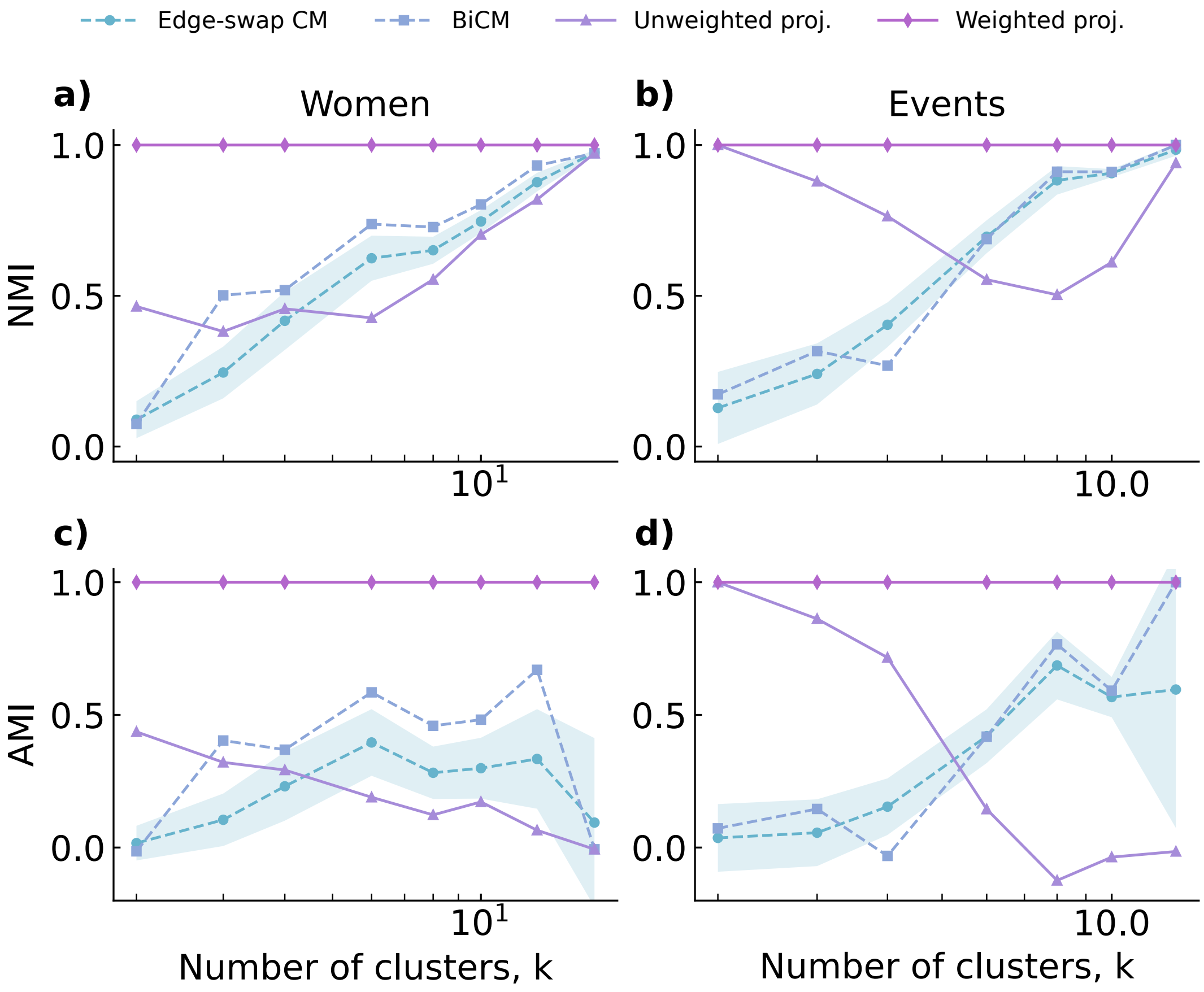


Figure 7. **Information-theoretic comparison for the Davis Southern Women network.** Normalized (NMI) and Adjusted Mutual Information (AMI) between the empirical bipartite hierarchy, null models, and one-mode projections. The weighted projection maintains high NMI and AMI across resolutions, indicating that local overlap captures most of the structural organization in this system.

### 3.2. Projection fidelity across complementary partitions

As discussed in the main text, one-mode projections distort bipartite geometry by collapsing multi-step alternating paths. To assess how this distortion depends on the underlying structure, we extend the NMI and AMI analysis to the complementary partitions of the datasets considered in Fig. 4: hosts in the host–virus network, and countries in the product–country and language–country systems (see Supplementary Fig. 8).

For the hosts (see Supplementary Fig. 8(a,d)), the results mirror those obtained for viruses. Degree-preserving null models (BiCM and edge-swap CM) yield AMI values close to zero across resolutions, indicating that the observed hierarchy cannot be explained by degree constraints alone. Projections, by contrast, retain a positive but incomplete similarity with the empirical dendrogram, showing that shared-neighbor overlap captures part of the organization while truncating longer-range correlations ($A-B-A-B$).

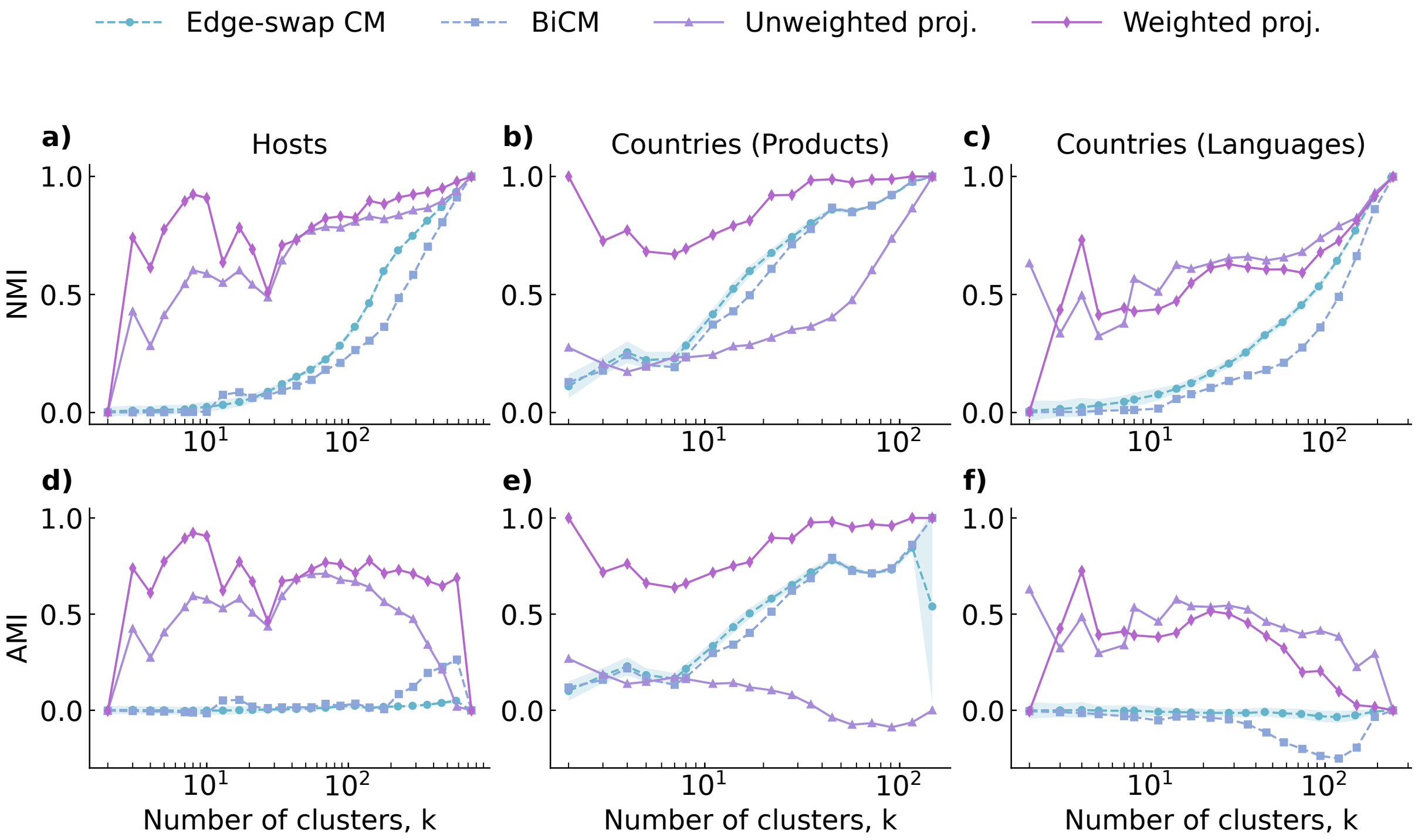


Figure 8. **Information-theoretic analysis of complementary partitions.** Comparison between clusterings obtained from the empirical bipartite dendrogram and those derived from degree-preserving bipartite null models (Edge-swap CM and BiCM) or from one-mode projections, as a function of the number of clusters $k$. **(a–c)** Normalized Mutual Information (NMI) for Hosts, Countries (Products), and Countries (Languages). The baseline agreement increases with $k$ for all models. **(d–f)** Adjusted Mutual Information (AMI), which removes chance agreement. Null models collapse to near-zero similarity, while projections retain a partial, resolution-dependent agreement. In the product space (e), weighted projections achieve higher similarity due to strong overlap redundancy.

The product–country network provides a contrasting regime. For the country hierarchy (see Supplementary Fig. 8(b,e)), the weighted projection shows high AMI values across a broad range of resolutions. This reflects the strong redundancy of export portfolios, which creates dense overlap patterns and makes short-range correlations dominant. In this regime, projection approximates the bipartite geometry. However, the unweighted projection fails to recover the structure, indicating that edge weights are essential to encode overlap intensity.

In the language–country network (see Supplementary Fig. 8(c,f)), projections recover only a limited fraction of the hierarchy, with AMI remaining moderate and strongly resolution-dependent.

Taken together, these results identify a structural criterion for projection validity: it is controlled by the extent to which multiscale organization is encoded in local overlap rather than in multi-step correlations. When a multiscale organization is dominated by short-range correlations, projections can approximate the bipartite structure; when longer-range dependencies are relevant, projections lead to systematic distortions.

## 3.3. Ecological interaction networks

### *3.3.1. Barrett plant–pollinator network: sparse, hierarchy-dominated regime*

As a representative ecological system, we consider the Barrett plant–pollinator network [1], a sparse mutualistic network with heterogeneous cross-class interactions of a boreal forest ecosystem. Unlike dense affiliation systems, its structure is not dominated by local overlap, but by longer-range dependencies across partitions.

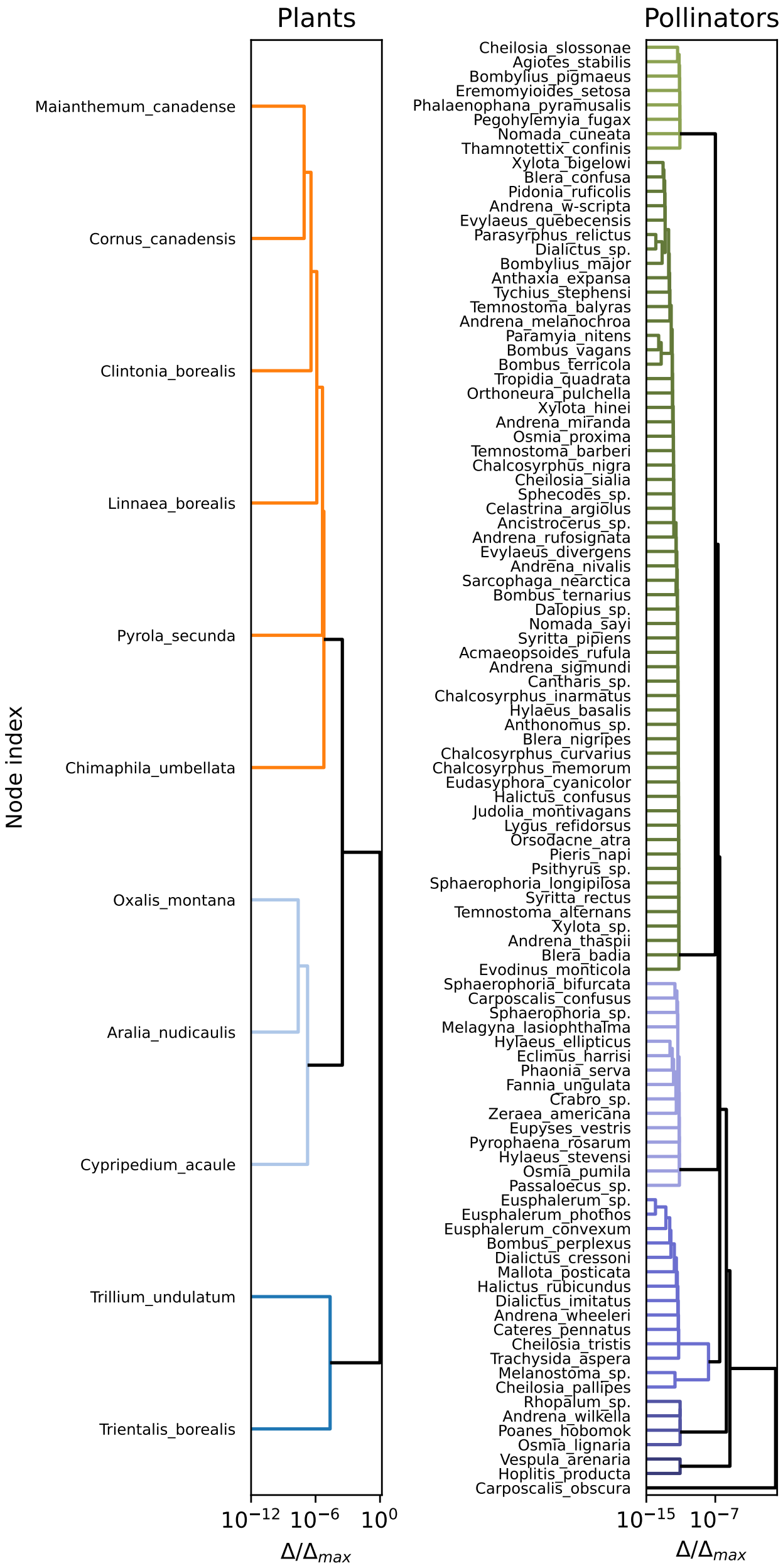


Figure 9. **Hierarchical organization of the Barrett network.** Bipartite diffusion dendrogram for plants and pollinators. This separates structurally specialized plants (e.g., *Cypripedium acaule*) into peripheral niches distinct from the generalist core. Similarly, diffusion-based hierarchy isolates a central core of highly connected generalist species (e.g., *Bombus* spp.) from specialized sub-lineages of solitary bees.

Applying the bipartite renormalization framework reveals a clear multiscale organization in both node classes (plants and pollinators, see Supplementary Fig. 9). The hierarchy is structured around a dense core of generalist species, which forms the backbone of the interaction network, together with peripheral branches corresponding to

more specialized niches.

The generalist core is populated by highly connected, ubiquitous pollinators, including bumblebees (e.g., *Bombus vagans*, *Bombus ternarius*) and widespread hoverflies. At finer scales, the hierarchy resolves more specialized sub-lineages, with peripheral clusters corresponding to narrowly interacting species, such as solitary bees (e.g., *Osmia lignaria*, *Hoplitis producta*) and structurally specialized plants (e.g., *Cypripedium acaule*).

This organization reflects a core–periphery structure in the interaction space, where generalists define the backbone of connectivity and specialists occupy distinct peripheral branches, highlighting the role of multi-step correlations in shaping ecological niches.

To quantify how this structure is affected by projection, we compare the bipartite hierarchy with degree-preserving null models and one-mode projections using NMI and AMI (see Supplementary Fig. 10).

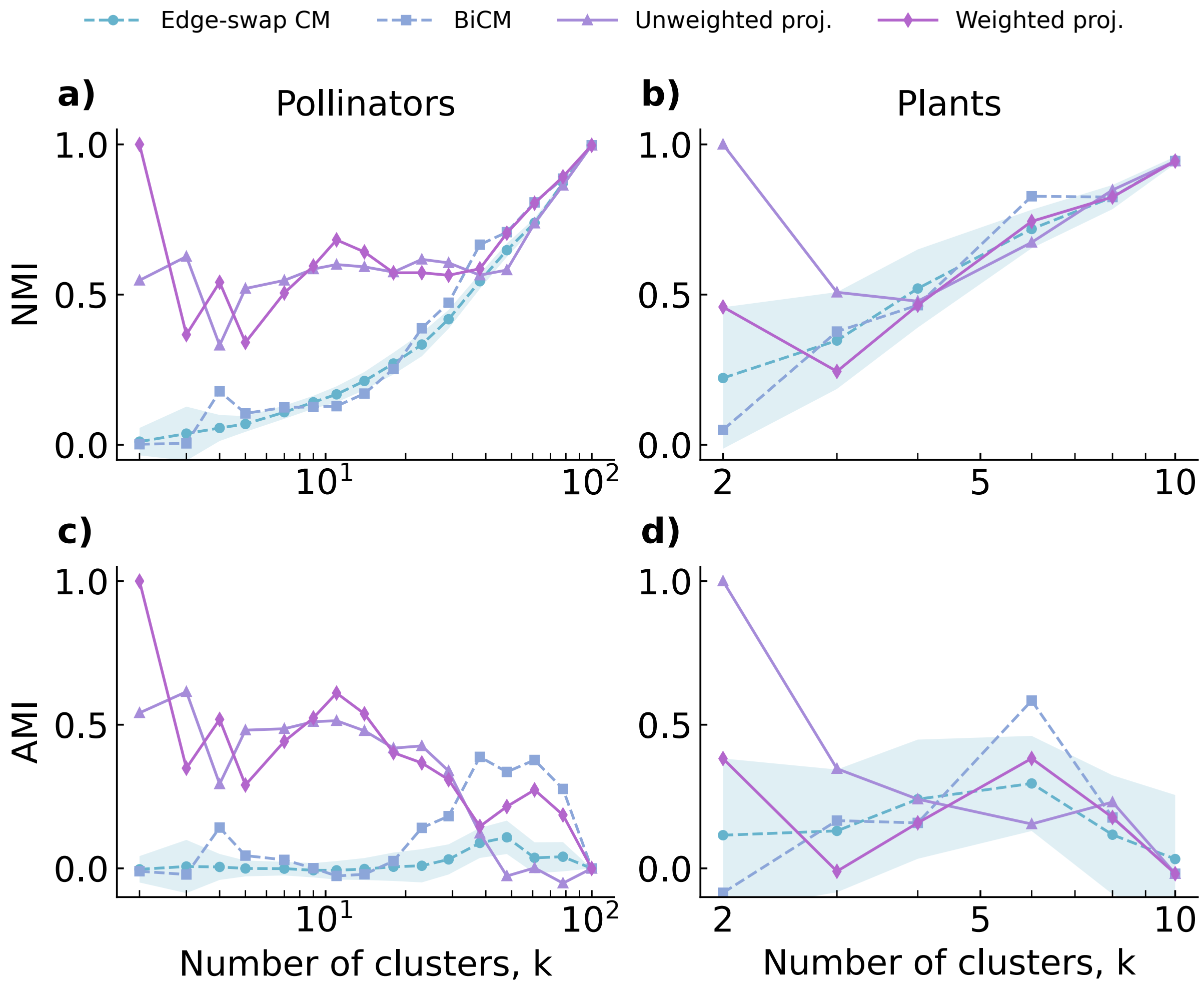


Figure 10. **Information-theoretic comparison for the Barrett network.** Normalized (NMI) and Adjusted Mutual Information (AMI) between the empirical bipartite hierarchy, null models, and one-mode projections. Null models collapse to near-zero AMI, while projections retain a partial and resolution-dependent similarity.

Under AMI, degree-preserving null models yield values close to zero across resolutions, indicating that the observed hierarchy cannot be explained by degree constraints alone. Projections retain a positive but incomplete similarity, capturing part of the local structure through shared-neighbor overlap while failing to reproduce the full hierarchy.

This behavior reflects the sparse and heterogeneous nature of the network: multiscale organization is driven by multi-step alternating paths rather than local density. As a result, projection truncates essential correlations and systematically distorts the diffusion geometry. This identifies a regime in which projection fails systematically, as multiscale organization is encoded in higher-order correlations beyond local overlap.

### 3.3.2. *Bartomeus plant–pollinator network: high-overlap and asymmetric projections*

We next consider a Mediterranean plant–pollinator network [2], characterized by a dense, highly nested core of generalist species (see Supplementary Fig. 11). This provides a complementary regime to the sparse Barrett network, allowing us to assess the impact of strong interaction redundancy on multiscale geometry.

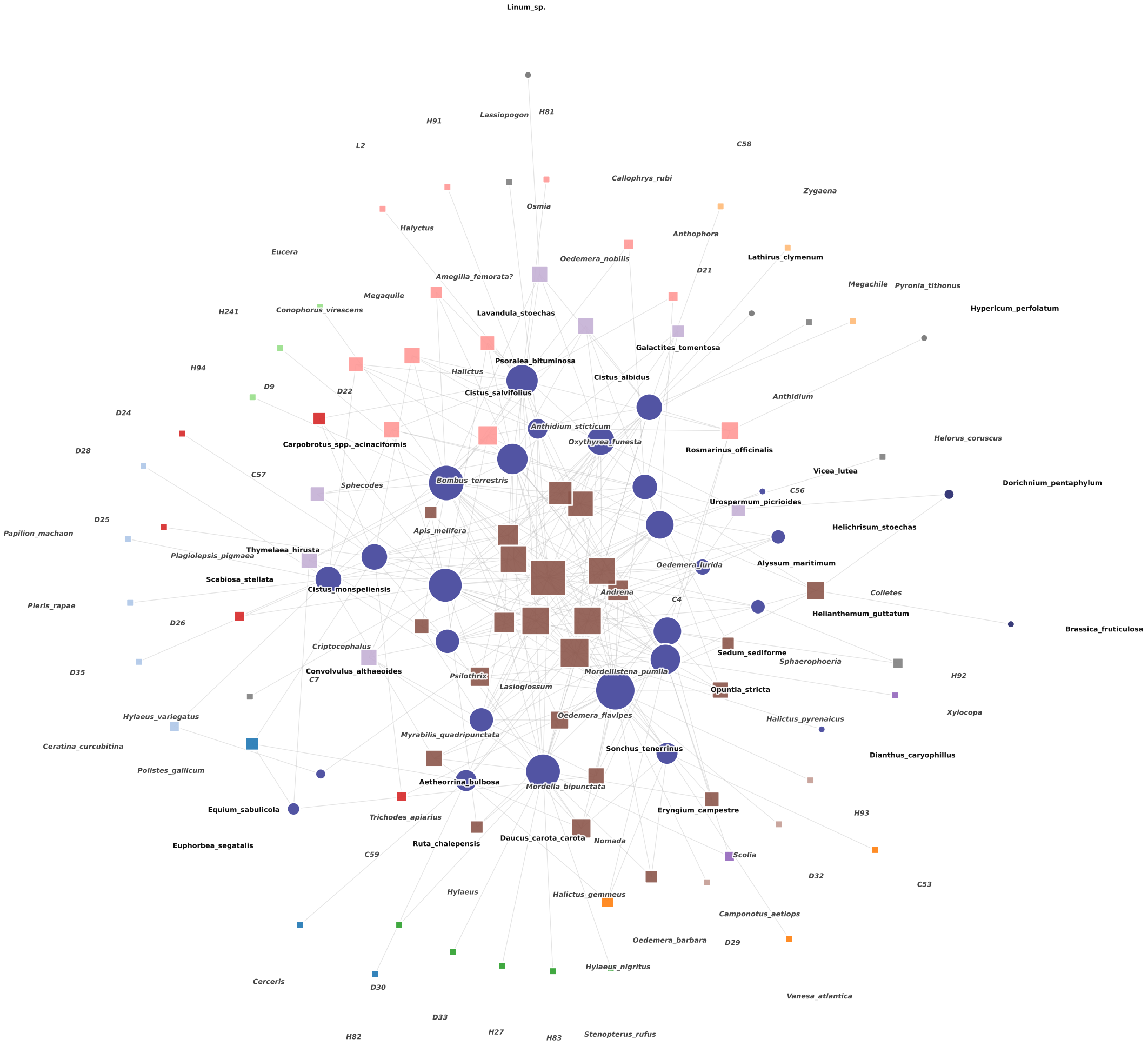


Figure 11. **Bipartite interaction network of the Bartomeus ecosystem.** The topology is dominated by a dense, nested core of highly connected generalist species, resulting in strong overlap redundancy across both node types.

The bipartite diffusion geometry reveals a core-dominated structure in both classes. The central module is populated by ubiquitous, highly connected species, including pollinators such as *Apis mellifera* and *Bombus terrestris*, while peripheral branches correspond to more specialized species (e.g., *Eucera*, *Polistes gallicus*). A similar pattern is observed for plants, where frequently visited species (e.g., *Cistus monspeliensis*, *Lavandula stoechas*) form a dense central module, distinct from more weakly connected taxa (e.g., *Linum*, *Brassica*).

This organization reflects a hierarchical core–periphery structure driven by highly redundant interaction patterns.

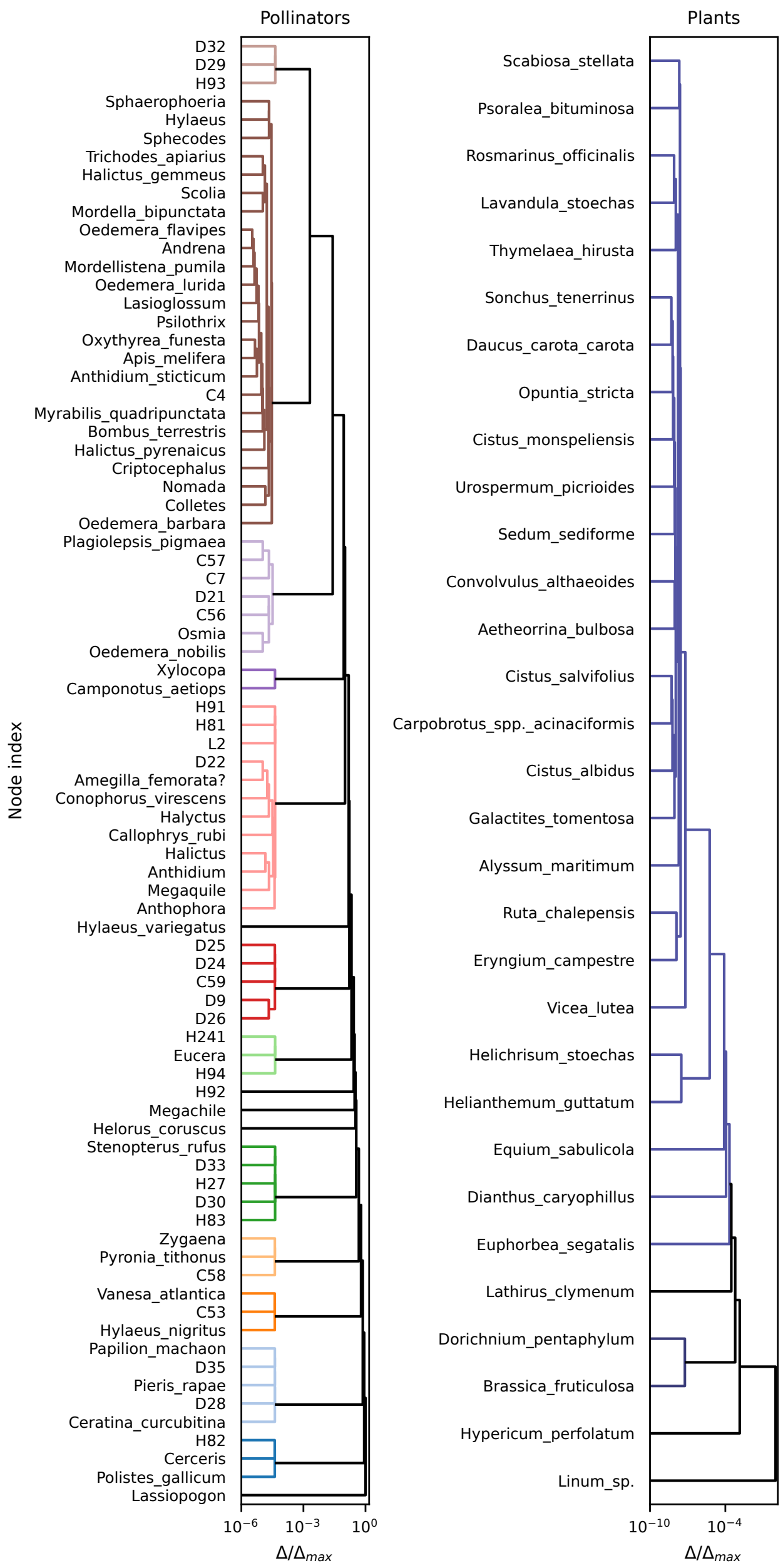


Figure 12. **Hierarchical organization of the Bartomeus network.** The diffusion-based hierarchy reveals a dense generalist core and peripheral branches associated with more specialized species, reflecting a strongly redundant interaction structure.

The information-theoretic comparison (reported in Supplementary Fig. 13) reveals a pronounced asymmetry between the two classes of nodes. For the pollinators, projections retain only a partial and resolution-dependent similarity with the bipartite hierarchy, while null models collapse to near-zero AMI.

By contrast, for the plants, the weighted projection achieves high AMI across a broad range of resolutions, approaching unity. This reflects the strong redundancy of interaction patterns: because many plant species share a highly overlapping set of generalist pollinators, the geometry is largely encoded in short-range correlations (paths of length two). In this regime, projection provides a good approximation of the bipartite structure.

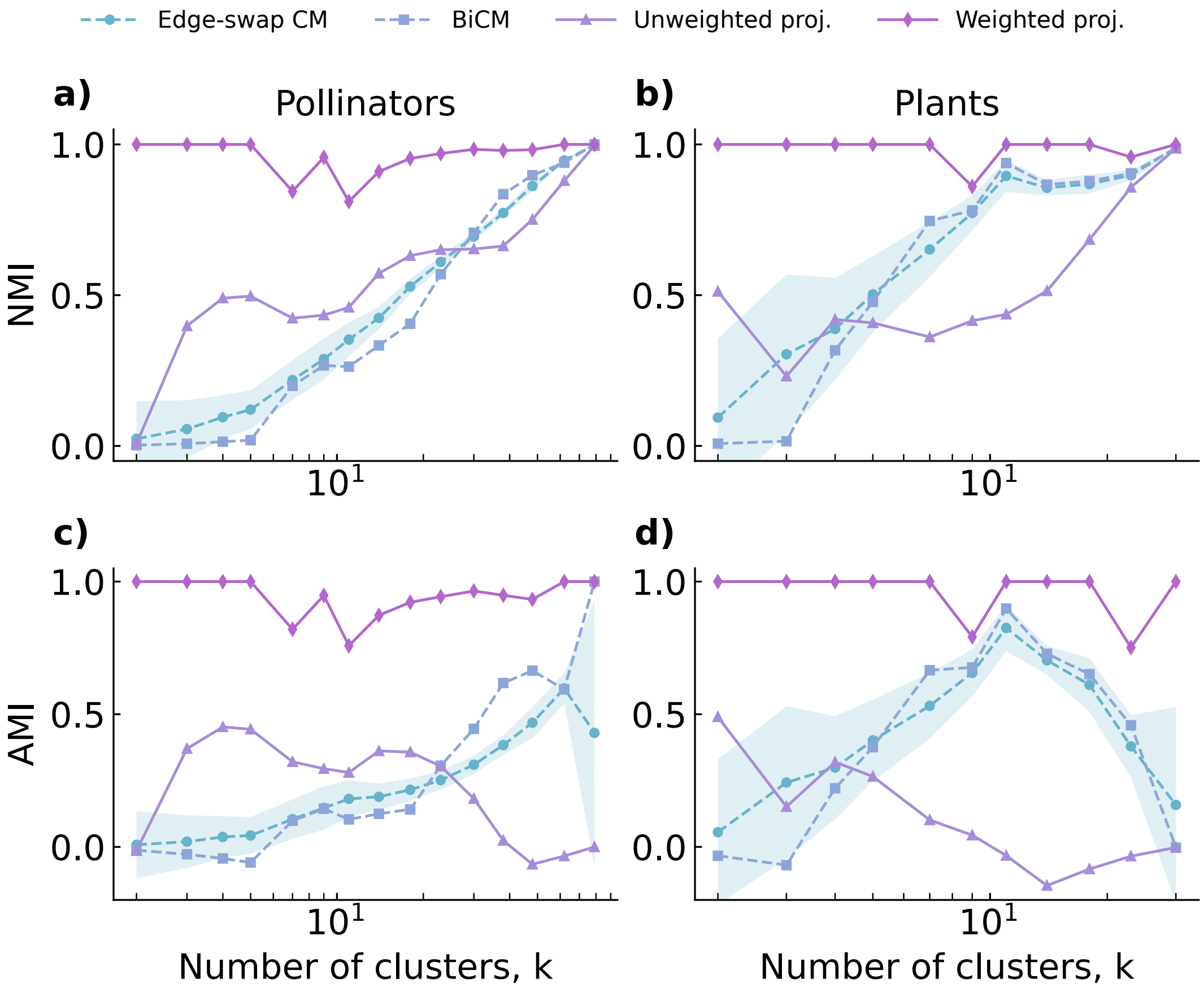


Figure 13. **Information-theoretic comparison for the Bartomeus plant-pollinator network.** NMI and AMI between the bipartite hierarchy, null models, and projections. The weighted projection closely approximates the plant hierarchy due to strong overlap redundancy, while the projection remains incomplete for the network of pollinators.

This asymmetry identifies a structural criterion for projection validity: it depends on whether multiscale organization is dominated by local overlap or by higher-order correlations. Different partitions of the same system may therefore exhibit qualitatively different projection behavior.

### 3.3.3. *Schleuning plant–frugivore network: core–periphery asymmetry*

To conclude our ecological analysis, we consider a tropical plant–frugivore network from a Kenyan forest [3], characterized by a diverse interaction space spanning avian and mammalian dispersers.

The bipartite diffusion geometry reveals a core–periphery organization in both node classes (see Supplementary Fig. 14). For dispersers, a dense generalist core emerges, comprising highly connected species such as *Pycnonotus barbatus*, *Turdus pelios*, and generalist primates (e.g., *Papio anubis*, *Cercopithecus ascanius*), while specialized species occupy peripheral branches. A similar large-scale structure is observed for plants, with a dominant core of frequently consumed species.

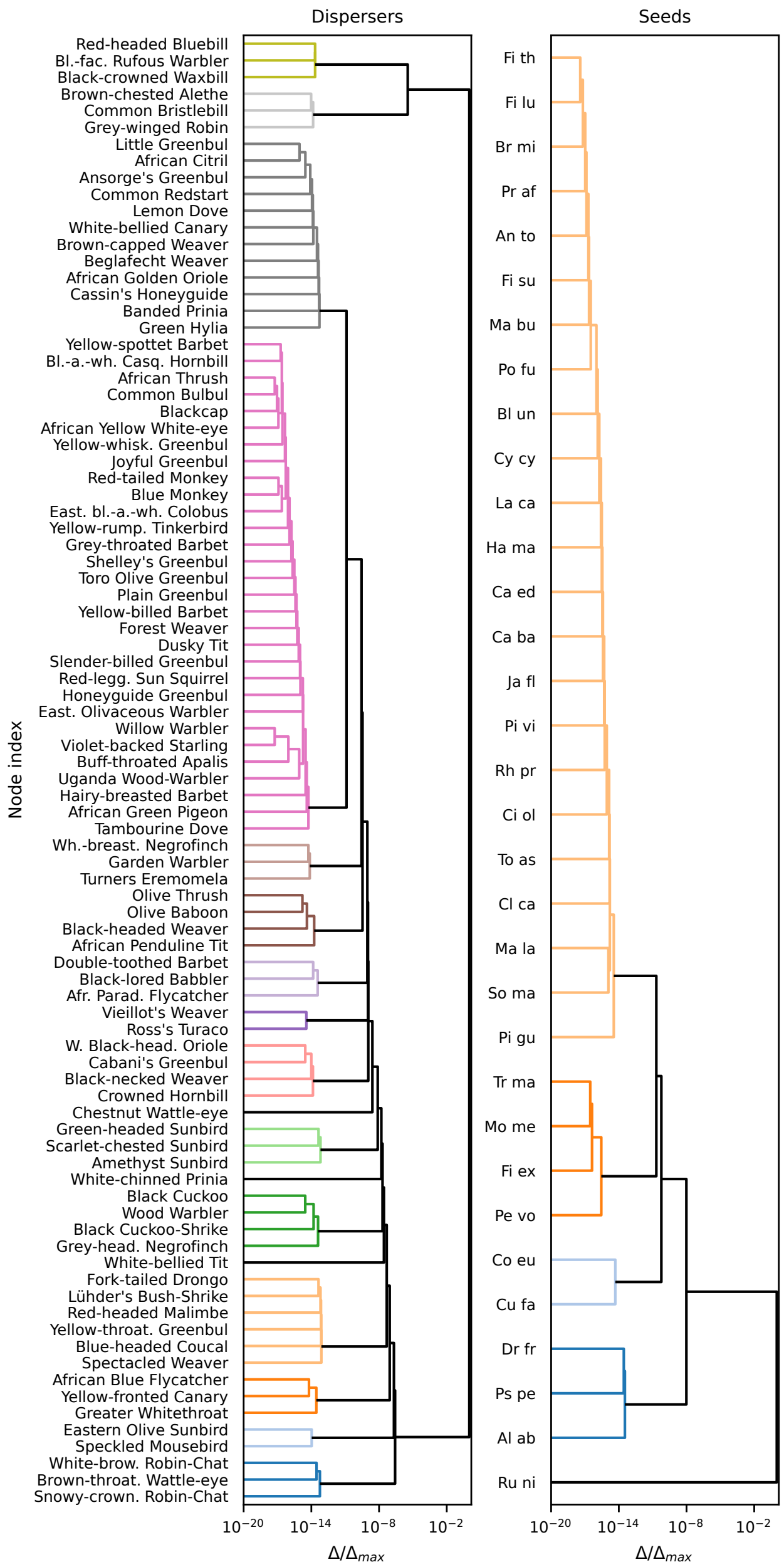


Figure 14. **Hierarchical organization of the Schleuning tropical plant–frugivore network.** The bipartite diffusion geometry reveals a massive, cohesive generalist core for both plants and frugivores. For dispersers, ubiquitous avian frugivores (e.g., Common Bulbul, African Thrush) cluster together with generalist primates (e.g., Olive Baboon, Red-tailed Monkey), forming the dense structural backbone of the ecosystem. Highly specialized species (e.g., Black-crowned Waxbill) are correctly isolated in peripheral, deep-branching lineages.

The information-theoretic analysis (see Supplementary Fig. 15) reveals a marked asymmetry between the two classes. For dispersers, weighted projections retain relatively high AMI across resolutions, reflecting the strong redundancy of shared diets and the dominance of short-range correlations. In contrast, for the seeds, projections fail to recover the hierarchy at intermediate and fine scales, indicating that such an organization depends on longer multi-step pathways ($A - B - A - B$) that are removed under projection.

At the finest resolutions, the AMI of degree-preserving null models increases for the seeds, showing that peripheral plant specialization is strongly constrained by the local degree sequence. This highlights a dual structural regime: a dense, redundancy-driven core and a degree-constrained periphery. The bipartite framework captures both consistently, whereas projection selectively preserves only the former.

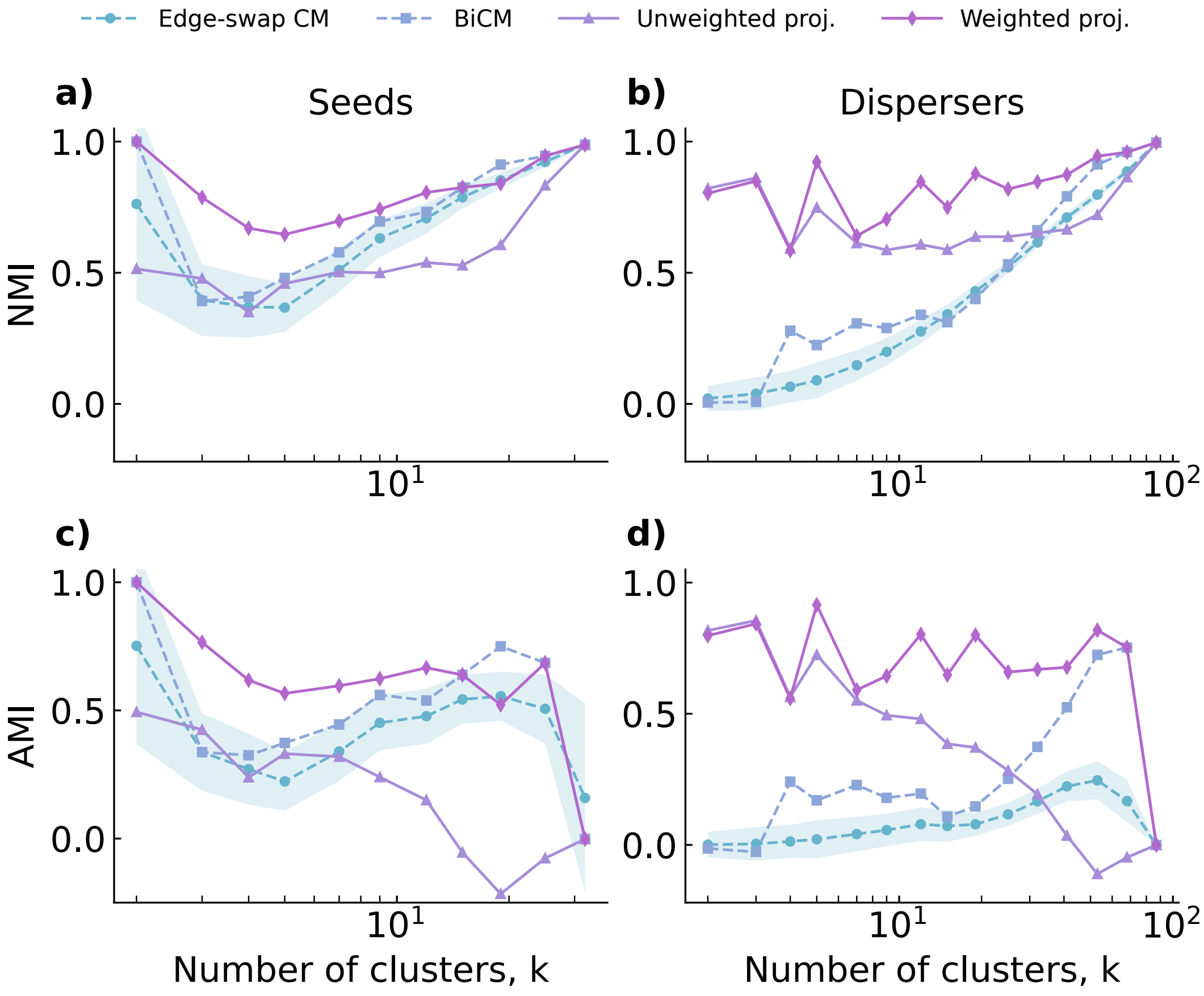


Figure 15. **AMI and NMI of the Schleuning network.** Normalized (NMI) and Adjusted Mutual Information (AMI) highlight the severe limitations of one-mode reductions. While the weighted projection captures the redundant structure of the dense disperser core (high AMI), it severely truncates the multiscale geometry of the seeds. Furthermore, the rising AMI of null models (BiCM, Edge-swap CM) at finer scales for the seeds reveals that peripheral plant specialization is heavily constrained by the local degree sequence—an essential structural feature that is entirely lost when the bipartite network is projected.

Our analyses reveal that the success of one-mode projections is fundamentally controlled by the geometry of correlations in the underlying bipartite system. When the organization is dominated by local overlap, projections retain a meaningful fraction of the structure. When instead it is sustained by multi-step alternating pathways, projection acts as a truncation of the diffusion geometry, erasing the correlations that define hierarchy across scales.

Importantly, this breakdown is not uniform: it depends on the empirical topology and can differ even between complementary partitions of the same network. The bipartite renormalization framework resolves this limitation by preserving the full interaction space, allowing multiscale organization to be consistently recovered across both dense and heterogeneous regimes.


\end{document}